# Trends of biosensing: plasmonics through miniaturization and quantum sensing

Giuseppina Simone

**Abstract**

Despite being extremely old concepts, plasmonics and surface plasmon resonance-based biosensors have been increasingly popular in the recent two decades due to the growing interest in nanooptics and are now of significant significance in regards to applications associated with human health. Plasmonics used in health surveillance systems have a high sensitivity and meets the standards for sensitivity, selectivity, and detection limit. In addition, integration into microsystems and point-of-care devices has enabled significant levels of sensitivity and limit of detection to be achieved. All of this has encouraged the expansion of the fields of study and market niches devoted to the creation of quick and incredibly sensitive label-free detection. The trend is reflected in the development of wearable plasmonic sensors as well as point-of-care applications for widespread applications, demonstrating the potential impact of the new generation of plasmonic biosensors on human well-being through the concepts of personalized medicine and global health. In this context, the aim here is to discuss the potential, limitations, and opportunities for improvement that have arisen as a result of the integration of plasmonics into microsystems and lab-on-chip over the past five years. Recent applications of plasmonic biosensors in microsystems and their sensor performance are analyzed. The final discussion focuses on the integration of microfluidics and lab-on-a-chip with quantum plasmonics technology as a promising solution for chemical and biological sensing applications. Aiming to overcome the limits given by quantum fluctuations and noise, research in the field of quantum plasmonic sensing for biological applications has flourished over the past decade. The significant advances in nanophotonics, plasmonics and microsystems used to create increasingly effective biosensors would continue to benefit this field if harnessed properly.

## *1. Introduction*

The pandemic events of the past two years have made scientists aware of the necessity to strengthen preventive and diagnostic safety plans owing to the conviction that rapid detection methods are primarily required to mitigate the risk. As a result, the new programs, methods, and timely control of any biological aggressiveness now depend heavily on early diagnosis. Emphasis has been placed on molecular detection research, which has the capacity to be a powerful tool to uncover the fundamental purposes of biological molecules and the mechanisms underlying various diseases. However, the performance standards for the detecting systems are tightening up and requiring ever-higher sensitivity levels. Sensitive systems are in demand because biological and clinical samples are typically only available in modest quantities, but even low concentrations of contaminants can induce infection. Specific sensors are needed to discriminate between analytes and components and to reduce the false positive rate. From the analysis of the current detection systems, it can be concluded that the transduction of analytical signals is still a challenge and very few physical methods are efficiently used for direct, real-time measurements.

Among others, optical biosensors, especially the plasmonic and photonic ones, have attracted much attention from scientists due to their many advantages, including their ease of use, high sensitivity, high specificity, and real-time capability without labeling single molecules [1] as well as the ability to run multiplex investigations [2,3]. For these reasons, they have found applications in many areas, such as in medical diagnosis [4], food safety and environmental monitoring (5–7). The flourishing development of nanophotonics and nanooptics over the past decade has supported the growth of plasmonic sensors and biosensors. Another impetus for the positive trend in sensors came from nano- and micro-technologies, which, by improving the manufacturing precision, have increasingly become a valuable tool in the hands of experts to provide prototypes and a platform on which the plasmonic optics in the sub-wavelength range can be examined. The simultaneous flourishing of nanooptics on the one hand and the development of raffinate nanotechologies on the other envision a trend that set the stage for the realization of very sensitive and reliable rapid detection systems; in fact, the integration of plasmonic biosensors into miniaturized systems overlaps with two leading technologies striving in the high-tech race for new areas of application and sources of continuous breakthroughs. Among the other advantages derived from this marriage, miniaturized and fluidic plasmonic biosensors are attractive because they allow real-time, label-free monitoring of molecular binding events, with interesting advantages in terms of cost reduction, high sensitivity and specificity, fast response and the possibility of multiplex detection [5,6] and have been identified as a promising area of development to contribute to diagnostic and clinical research. It offers a platform for molecular behavior in an aqueous environment using an ultra-strong plasmon confinement mechanism for in situ detection, offering the potential to implement nanoscale sensitivity induced by a reduction in the absorption of water molecules [5]. Moreover, the compatibility with other available technologies makes it possible to combine them with multiple sensing principles, which can improve the effectiveness of the overall measurement capabilities. An early example of this association showed that the integration of a plasmonic biosensor into a microfluidic device increased the ability to detect a short sequence of nucleic acids characteristic of *Escherichia coli* down to femtomole level in a few minutes [6]. Quantification of binding kinetics and events was proven to be routine; moreover, remarkable insights were gained for the detection of pathogens, in combination with other techniques [7] or optical characteristics [8].

Today, the majority of commercial plasmonic biosensors uses surface plasmon polaritons (SPPs) excitation in an attenuated total internal reflection configuration due to the consistency of the optomechanical configuration. However, this architecture seems slightly antagonist with the concept of miniaturization as it requires multiple offline optical components while other optomechanical configurations might perform better when integrated into small systems [9]. A relevant role is played by the localized surface plasmon polaritons that are excited in nanometric features allowing for sensitive analysis while maintaining a good level of portability. Not surprisingly, then, nanoparticles and nanofeatures with different shapes and compositions, as well as 2D assemblies (e.g arrays) have been produced and integrated into point-of-care systems to assess the efficacy with a crucial role that the design of the nanofeatures has on the performance of the system [10]. Besides, notable discoveries from ongoing research also include the design and fabrication of wearable plasmonic sensors [11], e.g. made of degradable and cheap materials, which comply with the incumbent environmental requirements. Moreover, this last generation of wearable sensors offers an example of personalized medicine through continuous monitoring of the fluids excreted by the body [12]. Progress in this direction is expected to provide a positive boost to a market that began in 1990 with the first commercialization of the surface plasmon resonance system, but that has also experienced a stalling due to factors such as cost and competitiveness. The actual trend shows that there is one more factor that will appeal to an alternative market scenario for this established technology. The reaching of a limit for the extension of the sensor's sensitivity and accuracy by the quantum fluctuations of light will mark the transition to a more refined method of plasmonic sensing [13]. Quantum metrology can make a significant contribution to improved sensitivity and precision beyond the signal-to-noise limit [14]. This theme has been further encouraged by recent advances in our understanding of the fundamental theory underlying quantum plasmonic systems as demonstrated by the development in quantum plasmonics recorded in recent years, which have enabled research teams to study the application of quantum metrology concepts to plasmonic sensors [13,15].

The exegesis (roadmap) of this overview is divided into three sections to make it understandable and useful for the readers. The fundamentals of plasmonic and plasmonic biosensing are presented in the first section to give a multidisciplinary overview and provide key concepts. In the second section, conventional plasmonic biosensing and integration with microfluidics are described. Advantages and problems related to the detection of integrated devices and devices used in research and industry are analyzed. The methods and ideas that have shifted conventional sensing from the classical to the quantum domain are described in section three, where some preliminary examples of quantum biosensing are presented. The review that photographs the state-of-the-art of plasmonic biosensors in the last five years is intended for committed scientists working in the broad areas of plasmonics, and the intended audience consists of multidisciplinary scientists with interests in biosensing, microfluidics, photonics and quantum optics.

## *2. Methods of Plasmonic Sensing*

### *2.1 Theory behind plasmonics*

The complex dielectric function $\varepsilon(\omega)$ (or complex permittivity), denoted for convenience in this manuscript as $\varepsilon_m$, which consists of a large and negative real component describing the plasmonic behavior and a small and positive imaginary component affecting the ohmic losses, can be used to treat the optical response of plasmonic materials, such as metal

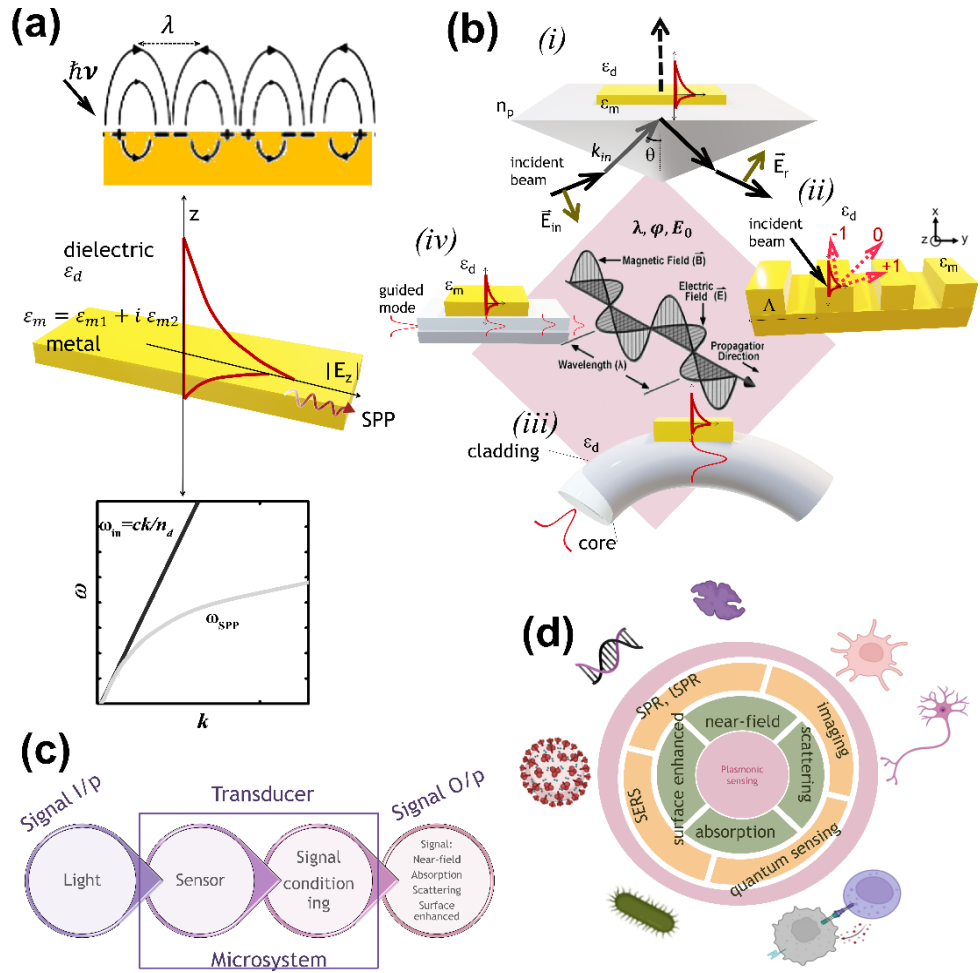

**Figure 1.** Plasmonic biosensor principle and integration in microfluidics. (a) Top: neutral plasma of the metal is polarized into positive and negative domains that produce a non-neutral charge distribution; middle: Evanescent field at the metal and dielectric interface; bottom: dispersion relation of surface plasmons compared to light in vacuum. (b) Surface plasmon and light coupling mechanisms: (i) prism, (ii) grating, (iii) fiber, (iv) waveguide. (c) Plasmonic sensor diagram. The light source comes in contact with the transducer which consists of a sensor and a signal conditioning element. Microfluidic integrated systems include the transducer. The output signal is recorded. (d) Plasmonic biosensing relies on several physical signals that give rise to some fundamental techniques for sensing biological samples (from RNA up to cell-cell interactions).

Upon exposure to light, the neutral plasma of the metal is polarized into positive and negative domains, creating a non-neutral charge distribution (Figure 1a, top). An electromagnetic wave outside the plasmonic material and an oscillating electron plasma into the metal are developed. The *quasi*-mode that couples the polarization of the electrons and propagates at the interface between the metal and the dielectric is defined as surface plasmon polariton, SPP and it represents a good example of light/matter coupling. The slow propagation of light in the metal is made possible by the nonlinear behavior of the dispersion produced by the solution of the Maxwell equations due to the slower propagation velocity at the metal/dielectric interface than in air. In turn, the two modes, that is, the electromagnetic wave in the dielectric medium and an oscillating electron plasma in the metal, both have an exponentially decaying evanescent behavior, but a distinct depth of field penetration in the two media, as visible by the distribution of the evanescent waves at the interface (Figure 1a, center). Figure 1a (bottom) shows the dispersion correlation under the condition of constant dielectric permittivity and complex metal permittivity. SPPs cannot be excited directly by light; in fact, it is shown that the light propagating in air (black line $\omega = c\, k/n_d$) cannot excite SPPs because $k_{spp} > \omega/c$, that is the wavevector for the SPP parallel to the surface is to the right of the light in air. The configuration does not allow the simultaneous achievement of energy and momentum conservation (($\omega_{in}(\omega) = \omega_{spp}(\omega),\ \ k_{in}(\omega) = k_{spp}(\omega)$)); in turn, the two conditions need to be established for the momentum transfer. Within technological applications, different approaches to the excitation of the surface plasmon polaritons have been investigated, which are shown schematically in Figure 1b. The coupling mechanisms can be divided into four groups: prismatic (Table (i)), grating (Table (ii)), fiber (Table (iii)) and waveguide ((Table (iv)) coupling. Fundamentals of surface plasmons and surface plasmon polaritons as given by the interaction between the photons and the plasmons play a crucial role in the design of the plasmonic sensor and remain the key concept of nanophotonic research, which has significant insights related to the tuning or transient control of properties is provided. Since plasmon resonance depends on the interaction of the light with the metal, it follows that tuning or momentarily controlling the light's wavevector, wavelength, amplitude, phase, and polarization allows tuning of the metal's plasmonic behavior, as determined by the equation of the electric field of the beam $E_x = E_0 cos(\omega t - kz + \varphi)$. All of this has sparked several experiments aimed at defining different strategies to get over the constraints imposed by the non-negligible loss that also characterizes noble metals. Many attempts have been generated for seeking new low-loss plasmonic materials with acceptable optical performance (e.g. doped-semiconductor, transparent conducting oxide, metal alloys, and graphene). These investigations have facilitated the extension of the list of plasmonic materials and expanded the fields of application. The fabrication of new structures of metamaterials that mimic the properties of plasmonic materials and SPP-like dispersion correlations has yielded artificial materials called spoof SPPs, which operate in microwave and terahertz frequency regimes as opposed to the typical frequencies of plasmons (visible and near-infrared) and have enabled a tunable and low-loss regime for practical plasmonic applications. Therefore, for example, plasmonic materials can be designed based on the structural dispersion of guided-wave propagation derived from the parallel-plate waveguides or rectangular waveguides that achieve negative permittivity capable of mimicking the property of plasmonic materials [16]. The technological approach chosen to design the system determines its overall performance, from plasmonic resonance to the sensitivity of the SPPs. Depending on the specific application, the device is designed by optimizing both the geometric features and the materials. For the latter, noble metals such as silver (Ag) and gold (Au) are the most interesting substrates for exciting SPPs; however, Au is preferred over Ag because of its chemical stability. Faster plasmonic mode propagation is imperative to obtain ultrathin gold films with low surface roughness and good adhesion to the substrate. This attention remains valid for all sensor systems based on metamaterials and hyperbolic materials, in which surface and bulk waves propagate with wavevectors considerably larger than those of light, and the thin active metal layer gives short-range surface plasmon propagation with a higher figure of merit.

## 2.2 Concept of biosensing and integration in microsystems

Figure 1c depicts a diagram of the plasmonic biosensor concept and its integration into microfluidic systems: the light and the metal transducer (e.g. thin film, nanoparticles, nanostructured surfaces, etc) interact and generate a signal that is able to perceive and translate the biological events. Techniques such as surface plasmon resonance (SPR), localized SPR, imaging, or surface-enhanced Raman scattering (SERS) are used to read the signal carrying the biological information (such as near-field, scattering, surface enhancement, or absorption) (Figure 1d). The sensitivity for detecting the targeted biological events depends on the characteristics of the transducer but also on how the sample is driven at the target area. Microsystem and microfluidic technologies enable the control and manipulation of liquids at the micron scale and have facilitated the development of total microanalysis systems or lab-on-a-chip. Moreover, microfluidic systems integrated with other detection technologies have enormous potential for use in clinical practice, as they enable a wide range of biological and biomedical experiments to be performed quickly and efficiently with a minimal amount of liquids and reagents. Furthermore, microfluidic devices are particularly efficient in the accurate and precise manipulation of biomolecules, cells, or tiny particles that is critical to various biological processes. The high level of development and automation that microsystems have recently achieved has significantly supported the progress of fully integrated biosensors. The intrinsic fluidic nature allows for rapid modification of the dielectric environment by injecting and mixing liquids with different dielectric properties. Sophisticated systems of valves and pumps apply tight control over sample delivery and activate a mechanism to toggle the activation of plasmonic features on and off. More specifically, from a fluid dynamics perspective, drag and Brownian motion balance the external forces and address the trajectory of the particles [17,18].

In the field of medicine, biosensing is considered a potent analytical tool for the early detection, diagnosis, and therapy of diseases. Enzymes, aptamers, immunological proteins such as antibodies and antigens, nucleic acids such as DNA and RNA, and microorganisms such as bacteria and pathogens, as well as cells are just some of the biomolecules that consider the early detection of cancer and benefits other diseases and that can be detected by miniaturized biosensors. Microsystems integrating plasmonics have gradually led to the design of increasingly complex architectures that have proliferated in both size and variety. After early research aimed at integrating fluid control methods with plasmonic biosensors, recent efforts have been made to realize all-in-one diagnostic devices at the point-of-care [19,20]. Indeed, plasmonic sensors have reached a level of sensitivity and detection limit comparable to full-scale sensors, motivating research into strategies for further miniaturization. Some coupling methods are exploited more than others to achieve reliable integration of the plasmonic transducer into a miniaturized and fluidic platform. Table 1 lists the main plasmonic biosensor concepts along with their advantages and possible disadvantages when applied in a fully integrated miniaturized environment.

| Sensor | Working principle | Microsystem | Advantages | Disadvantages |
| --- | --- | --- | --- | --- |
| SPR (planar sensor) | • prism coupling<br>• grating coupling | microchambers integrating single or multilayers | control of the dielectric environment by injecting and mixing liquids, sensitive demodulation | fluid dynamics, drag force and Brownian must be counterbalanced for the sampling control |
| Fiber | • grating coupling | 3D fluidic channels, fibers, multilayers | fully integrated system, flexibility, low cost, high-throughput investigations | sensitivity, reproducibility |
| Waveguide | • surface plasmon polaritons generated guided wave | 3D fluidic channels, fibers, multilayers | reduced facet reflection losses and simple engineering of index to convert the propagating modes | high precision of the alignments |
| localized surface plasmon resonance | • prism<br>• grating<br>• waveguide<br>• fiber | fluidic chamber | high control on the flow for enhancing the sample/optical feature contact time, 3D and 4D sample tracking | solid valving/pumping for sampling control |
| SPR imaging | • prism | multichannel | multiplexing, high throughput | channel cross-talk and signal contamination |
| Array | • see localized surface plasmon resonance | fluidic chambers or channels | high sensitivity leads by near-field concentration and confinement | reduction of absorption due to the water environment |
| Raman/SERS | • seelocalized surface plasmon resonance | fluidic chambers or channels | high control on the flow for enhancing the sample/optical feature contact time | expensive equipment, moderate integration |

**Table 1.** Pros/Cons of plasmonic biosensors integrated in microsystems

However, the merging of microfluidic modules and plasmonics has drawn some criticism, such as the difference in length scales, plasmonic measurements in nanometers and microfluidics in micrometers leads to the need to design the microfluidic/optical interface. The alignment between the input signal and the transducer, which is essential for the stimulation of the plasmon polaritons, and the detector of the output signal remains the most difficult operation at the microfluidic/optics interface. When the plasmon/photon interaction occurs via a feature used to modify the refractive index hierarchy (e.g. a prism), the alignment results are considerably challenging. In turn, plasmonic biosensing in microfluidics is an excellent fit with planar optical waveguides and Bragg reflectors acting as waveguides, resulting in a more compact configuration for miniature sensing devices than a prism-coupled approach [21]. Technologically, there are some other considerations, such as the microfluidic system must be optically transparent and based on simple and automated operations. While the first need can be easily met by choosing materials that are translucent and suitable for fabricating microfluidic devices, the operating systems, such as external tubing and pump parts, increase the complexity and size of the microfluidic platform. It is important to note that in this new level of integration of microfluidic functions, a significant improvement over their counterparts has been achieved. Typically, the process involves a significant change in the original configuration and/or the addition of external methods or materials, but also a simplification of the fluidic and optical architecture. From the plasmonic converter side, the integration path is smoother since most of the optical elements are designed in the sub-wavelength range, thus suitable for the miniaturization protocol.

## 2.3 Target samples for plasmonic fluidic biosensors

Multiple biological samples can be processed and analyzed by fully integrated sensors. The most common in vitro diagnostic methods use circulating proteins as the gold standard biomarker for disease identification and diagnosis. A dysfunction of cells, organs, or inflammatory processes is directly related to the overexpression, deregulation, or simply the presence of certain proteins in human tissues and fluids. Therefore, the rapid and accurate quantification of these biomolecules is crucial not only to detect a specific disease but also to determine its severity and prognosis. The high sensitivity required to identify minute amounts of protein and quantify them directly in complex clinical samples represents one of the main obstacles that plasmonic biosensors must address. Clinical investigations are feasible, and most importantly, the results of these investigations have also improved the understanding of multiple disease states, making it easier to compare the results of the analyses. A significant application of this approach has been developed for cancer diagnosis. Since cancer is the most common disease nowadays, most applications focus on early detection of the disease. For example, a lab-on-a-chip platform has been used to detect prostate-specific antigens from cancer patients with excellent accuracy [22]. For instance, antibodies present in the early stages in the serum of colon cancer patients were detected by a nanoparticle-enhanced SPR imaging platform [23]. Another common condition affecting the majority of the population is inflammation, which can also result from a variety of cancers. The diagnosis in this case can be made by identifying the dysregulation of various types of cytokines in the blood. In addition, nanoplasmonics technology may potentially be useful in treating autoimmune and neurological disorders, including Alzheimer's disease [16], which can be targeted by targeting the tau protein. While proteomics paved the way for early detection, the nucleic acid is now emerging as the leading competitive biomarker for early detection, prognosis, and evaluating the effectiveness of therapies in complex diseases. The origin of several important diseases, such as cancer, has been linked to gradually accumulating genetic mutations. Understanding the process behind epigenetics, which involves DNA methylation, microRNAs, and the regulation of mRNA, is key to early cancer prediction. Therefore, studying epigenetic mechanisms has helped researchers gain a thorough understanding of the different pathways cancer cells use to survive and proliferate normal cells, as well as to develop new effective therapies. Nucleic acid detection has found a potential platform in plasmonic and nanoplasmonic biosensors. However, difficulties can arise when using nucleic acids as biomarkers due to their low concentration and sequence similarities, which may be close to the limit of mismatching. Integration with microfluidics represents an approach to overcome the bottleneck as it gives the plasmonic sensor the potential to reach attomolar concentrations [20]. The third population of samples that biosensors can detect includes whole cells, single cells, and pathogens. Infections are typically caused by invading pathogenic organisms, such as bacteria or viruses, which multiply rapidly and produce toxins that elicit an immune response. Pathogenic infections can spread rapidly among humans, infecting entire communities and causing epidemics. To reduce the significant burden of infectious diseases worldwide, sensitive, selective and early identification of pathogens is therefore essential. Fluidic plasmonic biosensors represent an important platform to capture and analyze the samples despite some critical problems given by the large size of the samples and the fluidic mass transport [19]. The nanoscale dielectric environment of the metallic host structure has a significant impact on the efficiency of these plasmonic phenomena and opens easy avenues for optical biosensing at the molecular level and living cells, for which is crucial the control of the surrounding dielectric conditions.

## 3. Spectroscopic investigations in the microsystem environment

### 3.1 Plasmonic microsystems based on the prism and grating coupling

Coupling through a prism is developed by well-established optomechanical configurations (e.g. Kretschmann Figure 1b(i), Otto) operating under the conditions of total internal reflection. The resonant excitation of the SPP wave propagating at the metal/dielectric interface occurs at a unique angle and wavelength at which the energy and momentum matching condition between photon and surface plasmon occurs. In a dispersion diagram (Figure 2a), the crossing between the line of the light and the surface plasmon occurs through a prism because the wavevector of the light is affected by the refractive index of the prism as well as by the angle of incidence ($\omega = c\, k/n_p \sin\vartheta$), and under certain conditions of incidence, the SPPs can be excited. A decrease in the reflectivity signal measured in accordance with the incident angle or the excitation's wavelength can be used to detect the resonant excitation of SPPs. Additionally, when an analyte is deposited at the interface with the film metal, the resonance is excited via so-called evanescent field coupling, while the change in light intensity or shift in transmitted excitation light wavelength caused by the interaction of the analyte's molecules and the evanescent field is related to the change in refractive index (Figure 2b). The quantification of the shift enables the detection of the changes in the permittivity $\varepsilon_a$ of the biomolecules in contact with the plasmonic sensor. The sensitivity to the signal dip allows the estimation of various types of implicit parameters characterizing the optical properties of the analyte as well as the detection of chemical and biological binding events. Therefore, the prism-coupled biosensors are successfully used to characterize kinetic parameters, such as the equilibrium constant, dissociation, and association constants. Furthermore, due to the variety of fabrication methods ranging from the sandwich of different layers (Figure 2c) to more complex structures realized by precise photolithographic techniques, multilayers designed to merge the properties of different coatings from antireflective to SPP transparent have been exploited for developing diagnostic devices. The main benefit of using a microfluidic platform is related to the potential to create a highly controlled environment that allows for the binding activity of molecules such as antigens/antibodies [24], protein/protein, and protein/glycan [25]. Galactose/lectin binding analysis is shown in Figure 2d; the constant of dissociation $K_D$ is provided together with the change in signal strength, which is recorded by increasing or lowering the ligand concentration (panel i) and the experimental limit of detection of the system (panel ii). Additionally, due to the microsystem's low susceptibility to environmental factors, it is possible to precisely analyze the binding events and provide a model of the binding mechanism (panel iii). A similar protocol has been exploited in the analysis with anti-reflective layers alternating with $TiO_2$ and $SiO_2$, as well as with perovskite materials e.g. $BaTiO_3$, $PbTiO_3$, and $SrTiO_3$ have achieved an improved sensitivity in the detection of SARS-CoV-2 antigens compared to Au, resulting in a maximum value of 392 deg $RIU^{-1}$ [26]. Increasing the manufacturing precision to an atomically thin layer, the phase singularity achieved allows a lateral position shift and offers the advantage of increasing the sensitivity at a femtomolar level, even of lightweight molecules, such as the TNF-α cancer marker [27]. As a surface-based technique, plasmonic sensors can only detect molecules near the surface; therefore, the diffusion limit represents another problem that needs to be solved in classical SPR sensors, in order to enhance their performance. The Kretschmann design of the SPR sensor offers a workaround by allowing different sensing methods to be mixed and combined in efficient microsystems. For instance, the target analytes can be caged onto the detecting surface at the time of measurement using AC electrokinetic effects [28]; the diffusion limit related to the small particles in a fluidic environment can be overcome together with improved efficiency of collection capacities for objects of various sizes.

The grating-coupling principle is shown in Figure 1b (ii); the plasmonic behavior occurs at the interface with a metallic surface modulated by periodic corrugation [29]. The light/grating coupling shown in Figure 2e leads to a Bloch condition that depends on the geometric features of the surface (e.g. the grating period $\Lambda$). The wavevector relationship $k_{x,m} = k_x + 2\pi m/\Lambda$ (m: diffraction order, -1,0,1) emphasizes the in-plane wavevector that can scale up with the grating factor $2\pi m/\Lambda$. The latter condition allows the light dispersion to be shifted and coupled to the SPP dispersion, and even excites multiple surface polaritons. Plasmonic gratings represent a successful example of applying plasmonic sensing in a point-of-care platform. Integration with microfluidic chambers has simplified the detection procedures and limited the injected volume, providing a strong contribution to sensitivity enhancement. Figure 2f displays an example of a microfluidic system (panel i) where the active area consists of a uniform surface patterns (panel ii) present on optical disc-based metasurfaces to excite asymmetric plasmonic modes. The system enables a tunable Fano optical resonance in a microfluidic channel used for multiple target detection in the visible wavelength range. SARS-CoV-2 was detected with high sensitivity and specificity, which enabled the distinction of the pandemic virus from the influenza virus (panel iii) [30]. Most grating-based couplers use the phase method [31] or the combination of the sensing and imaging capabilities of guided-mode resonance devices, in which the grated substrate is embedded into a microfluidic chip. Investigations have been made possible by tracking the signal differences caused by changes in the diffraction pattern when the molecules are attached to the surface. With a detection limit of a few pM, the properties of the first-order diffraction pattern of a convex holographic grating were studied for detecting molecular activities, including biomolecular interactions with anti-human immunoglobulin G and human immunoglobulin G.

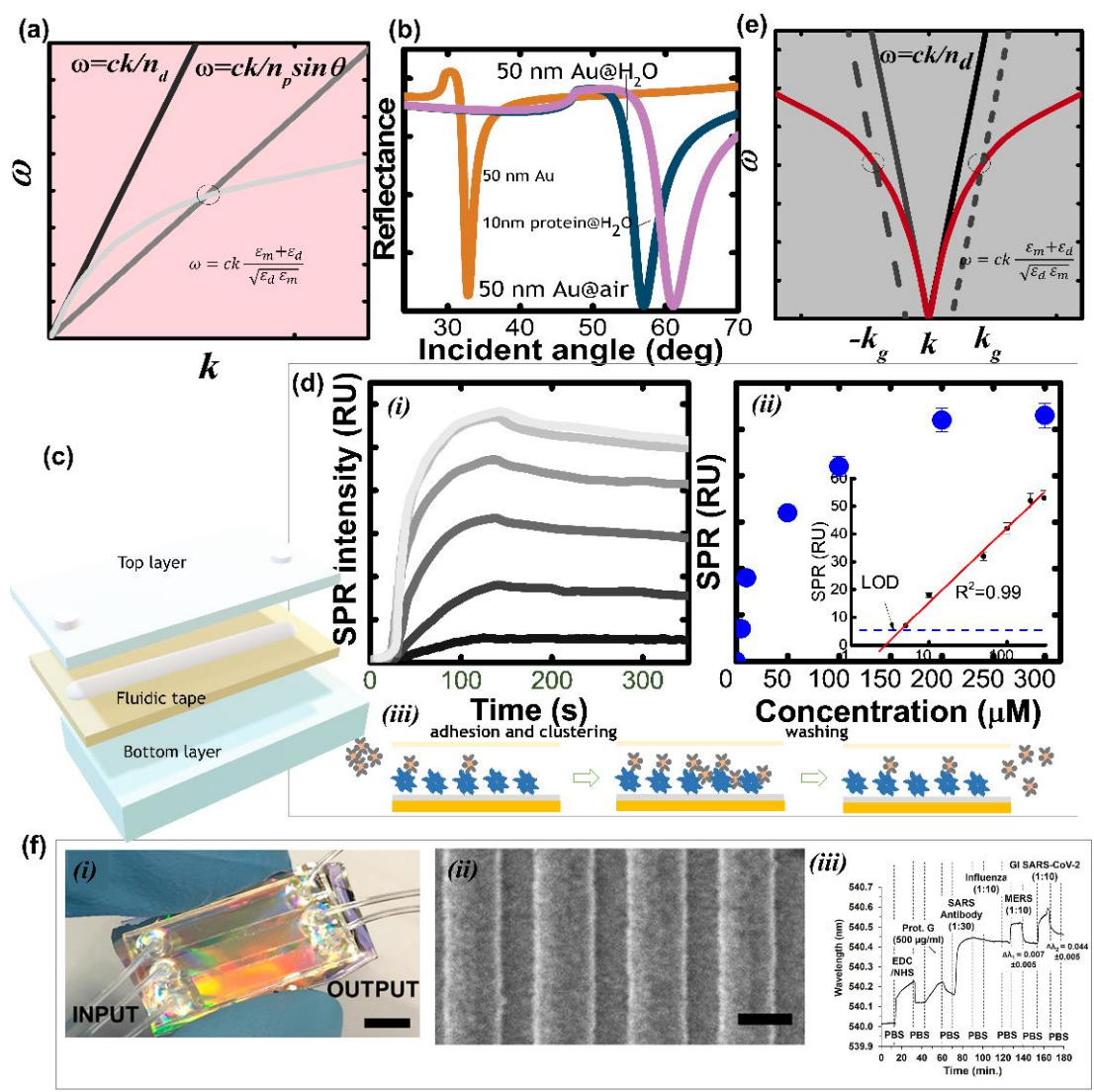

**Figure 2.** Prism and grating coupling. (a) Dispersion diagram showing the surface plasmon and the light coupling. (b) Reflectance in air and $H_2O$ at $\lambda$=650 nm of a 50 nm Au layer; the violet curve on the right side of the diagram shows the shift following the adsorption of a protein layer. (c) Schematic of a multilayer fluidic chip used for studying the binding kinetics of glycan/lectin system. (d) Kinetic analysis: (i) The sensorgrams are relative to the concentrations ranging from 5 mM up to 300 mM of lectin injected into the microfluidic channel; (ii) signal according to the concentration and (iii) binding activity mode including the steps of adhesion and clustering. (e) Dispersion diagram in grating coupling conditions displaying the influence of the order of grating. (f) Grating sensing for viruse detection: (i) plasmonic metasurface-based microfluidic chip f. Scale bar = 2.0 cm; (ii) High-resolution micrograph of the surface morphology of a plastic-templated DVD metasurface; (iii) Sensorgram showing the differentiation among viruses (influenza, MERS, and SARS-CoV-2). Copyright © 2022, American Chemical Society [30].

Grated-based sensors have also the advantage to offer an active environment for plasmonic imaging, a method of sensing that will be discussed later, however here it is worth noting that by performing molecular detection through imaging and diffraction pattern analysis, the method based on the use of a microfluidic system and the opto-mechanical design enables a label- and spectrometer-free detection [32]. The flexibility of the system can easily be extended to track multiple analytes as well as biomarkers [33]. In some cases, the grating is generated by fiber embedding Bragg grating surfaces, which may be coated with metals such as Au and Ag. These systems can be incorporated into a microfluidic environment, designed to regulate flow rates and ensure sample stability. The system is optically flexible thanks to the microfluidic environment, and indeed a twenty times more sensitive demodulation technique can be followed by cutting the upper and lower envelopes of metal-coated grating spectra. A similar approach has been followed for detecting cancer biomarkers, such as the Human Epidermal Growth Factor Receptor-2 proteins [34]. Furthermore, gratings were combined with different detection techniques to study the magnitude of the increase in sensitivity, and also to achieve improved sample handling. A relevant example is the magneto-optical surface plasmon resonance sensors, which recently have achieved a figure-of-merit as high as about $10^2$. In turn, with a sensitivity of about 100 RIU deg$^{-1}$, label-free and parallel detection of several disease-specific biomarkers, such as troponin, procalcitonin, and C-Reactive Protein, in undiluted urine samples, was obtained. The result was reached by combining the sensing and imaging capabilities of a chirped guided-mode resonance, obtaining low detection limits of 10 pg mL$^{-1}$ for all proteins [35]. Under this circumstance, the sensor is based on the transverse magneto-optical Kerr effect in a ferromagnet coupled with a noble-metal grating, which consists of a subwavelength periodic gold grating configured as a magneto-plasmonic heterostructure. Sharp, high-amplitude Fano-like magneto-optical signals can be obtained and used to detect very small changes in the refractive index of an analyte surrounding the sensor.

### *3.2 Plasmonic microsystems based on waveguide and fiber coupling through a mechanism of light squeezing*

Aside from triggering the SPPs, the nonlinear dispersion of the metals has shown an interesting response to small dielectric gaps between metal layers. The experiment was conducted in 2006 and demonstrated the possibility of squeezing light into a sub-wavelength-sized gap [36]. A similar configuration agrees that while frequency is unaffected, the wave shrinks significantly more than when traveling in free space and stores the same wavelength as it enters and exits the plasmon gap. The waveguide mode and localized SPR are generated by a mechanism of light squeezing, and the microfluidic environment creates a perfect environment for housing waveguides and nanometric features, consequently, this integration offers some advantages to the performance of the system [37][38][39][40]. It is worth noting at this point that some of the aspects of waveguide and fiber coupling related to light squeezing are considered to explain the quantum enhancement in plasmonic biosensing.

The design of plasmonic biosensors based on waveguide coupling entails a transistor-like component where light enters and exits, and the optical mode is controlled when inside the transistor (Figure 1b (iv)) [41]. The metallic structure concentrates the light to a subwavelength volume, which gives rise to a strong electric field having a modulus resulting in the following equation $E = E_0 \frac{(1+\kappa)\varepsilon_m}{\varepsilon_w + \kappa\varepsilon_m}$, which highlights dependence on the permittivity of the metal $\varepsilon_m$ as well as on the permittivity of the waveguide $\varepsilon_w$. The numerical solution of the components of the electric and magnetic field confinement at the waveguide's margins, where they are noticeably enhanced in comparison to the electric field of the exciting beam, is shown in Figure 3a. The distribution of the electromagnetic field, as well as the confinement inside the narrow volume of the waveguide, represents a remarkable advantage for the analysis of the biological signal and it represents a unique opportunity to study the activity of organisms and living cells in particular when combined with other techniques such as surface-enhanced Raman scattering; moreover, an important role in the mechanism of analysis is played by the material selected for the optical system fabrication as well [42] [43]. Several key findings were reported, and the most important are summarized here. An enhanced refractive index measurement was achieved by exploiting the singular phase of light associated with Brewster modes excited in a microfluidic hyperbolic metamaterial sensor composed of a multilayer stack of TiN (as plasmonic material) and $Sb_2S_3$ used as the phase change material [44]. By exploiting the potential of these systems and monitoring the propagation of the red blood cells, it was possible to measure pH fluctuations with an accuracy of 95% and to diagnose associated blood disorders. To improve the efficiency of the waveguide sensing mechanism, one strategy is to combine the transducer element including the waveguide with distinctive optical or chemical characteristics. More complex sensors, including combined plasmonic-electrochemical sensing within a microfluidic platform, achieve the sensitivity to probe proteins with a detection limit for mass measurement of $4×10^{-6}$ RIU. By including an electrochemical sensor, the sensing response associated with the waveguide transduction of the signal resulting from the injection of fluids at different concentrations into the microfluidic chip can be improved. Therefore, Au strips as electrodes with nearby Pt strips integrated as on-chip counter electrodes have been shown to reduce the standard stepped mass readout for the waveguide biosensors to values equal to 0.13 μW for the variation of the refractive index of protein solution $\Delta n = 5×10^{-4}$ RIU [45].

The fiber coupling mechanism is shown in Figure 1b(iii). The optical fibers are designed to guide light with low loss through a total internal reflection mechanism guaranteed by a core/cladding geometry and by the relationship among the material refractive indices. The light travels into the core, while the cladding prevents it from reaching the surroundings. The excitation of the plasmonic resonance is achieved by making contact between the plasmonic layer and the fiber core. In some cases, the coupling between the light and the core modes occurs through a forward-propagating cladding mode. However, the coupling can occur via a mechanism of reflection of the core field directly onto the cladding/sample interface through a tilted grating surface that reproduces a mechanism analogous to the prism coupling. Cladding mode can be excited by depositing the plasmonic metal on the fiber cladding; the plasmon resonance of the metal and the cladding mode match under phase conditions and perturb the refractive index of the cladding. Advantages such as flexibility and low cost make the fiber-based plasmonic sensors interesting for applications related to widespread use (viral infections, detection of toxins in water and food). Based on their configuration which is characterized by a huge extent of flexibility, fiber-based sensors lend themselves to be excellent for realizing wearable sensing devices that can achieve a thorough understanding of a person's physiology through the continuous and non-invasive monitoring of biochemical data. Furthermore, when combined to support tailored treatment modalities, wearable diagnostic platforms can play an essential role in preventing infections and diseases. An outstanding example is represented by the one-step optical core-shell microfiber textiles with single-walled carbon nanotubes. An interesting application is shown in Figure 3b where the fiber (panel i) is made of a hydrogel with high flexibility and a unique value of permeability that can provide an elevated sensitivity for the molecular detection of SARS-CoV-2 when engineered with plasmonic nanoparticles [46]. The 3D structure provided by a gel confers to the system the possibility to study the interaction between light and living cells (panel ii), for quantifying and digitalizing complex biological phenomena, such as 3D cancer progression and drug susceptibility (panel iii). The fibers' ability to provide real-time optical monitoring was used to measure hydrogen peroxide levels in wounds that exhibit a free decay signal over time. The system includes carbon nanotubes that enhance the strength of the device because of the multiple chiralities of the carbon nanotubes, which produce narrow-bandwidth near-infrared fluorescence that enables ratiometric signal readout independent of excitation source distance and exposure time [47].

**3.3 *Localized surface plasmon resonance microsystems***

The ability to tightly confine light has motivated the development of nanometric structures that control the optical mode within them. In the case of metallic objects with a dimension smaller than the wavelength of light, the delocalized charge carriers induce vibration rather than propagation of the electrons; Figure 3c illustrates the formation of positive and negative domains at the metal-dielectric surface. At plasmon resonance, the nanometer structures achieve strong resonant absorption of the light from the metal's polarization response, and the electrons within the metal particle produce a resonant susceptibility similar to what they would do if polarized in an electric field. While the resonant behavior of the particles results from the imaginary part of the susceptibility due to a phase shift, the real part of the susceptibility, which arises from absorption or scattering, enhances the plasmonic behavior and improves both the sensitivity and the tunability of the resonance. There are significant implications for sensor applications aiming to increase the yield of the sensor (e.g. sensitivity) to as low as the single-molecule detection limit by modifying the dispersion properties of the light and tuning the wavelength [48] [49,50] to the functionality of the design up to the final application, the shape, dimensions, and materials of the nano-elements open up for very specific bio-analyses. For the sake of completeness, it should be noted that the selection of the molecules required for the functionalization of the sensor, which is necessary to establish precise and selective bindings with the target analytes, leads to an additional degree of design freedom for the plasmonic device. The incorporation of nanoparticles and nanofeatures into microfluidics has been shown to have multiple advantages for studying biological samples and for converting 3D data acquisition into 4D by monitoring the biomolecular binding events in real-time. However, the flow and streamlines in the surrounding region must be considered when constructing the microfluidic environment that integrates optical nanofeatures for sensors based on localized surface plasmon resonance more than for the other plasmonic layouts. In fact, despite the sensitivity reaching a single-molecule level, the applicability of nanofeatures in a fluid environment also depends on the Brownian diffusion of the biomolecule to the sensor surface that dictates then the overall duration of the sensor assay. It follows that the design of the plasmonic nanofeatures and the microfluidic environment must optimize the sensing activity. Besides, it is interesting to note that the optical nanofeatures and the microfluidic environment can mimic the extra and intracellular environement, therefore it becomes more and more important to aim at a well-designed environment for living cell applications. Metallic nanoparticles of various shapes and compositions have been developed for integration into microfluidic devices, exploiting localized surface plasmon polaritons, and detecting the molecular activity of biological analytes such as the exosome and cytokine storm for SARS-CoV-2 diagnosis [51]. This latter application has been successfully demonstrated using a system with 100 nm Ag cubes in a digital microfluidic system [52]. In this context, the increased attention to environmental aspects aimed at reducing the impact on nature has led to the reconversion of industrial by-products. A prominent example is the recycled mill scale (e.g. iron oxide pigments) that has been converted into valuable products, in particular, many nanometric plasmonic structures [53] *In situ* and label-free high-throughput and multiplex cytokine secretion from adipose tissue have been demonstrated in a fully integrated system based on localized surface plasmon resonance [54]. A cartoon of the microfluidic platform is shown in Figure 3d(i), while an example of Au/Ag nanofeatures for exciting local surface plasmon polariton is shown in Figure 3d(ii) [55].

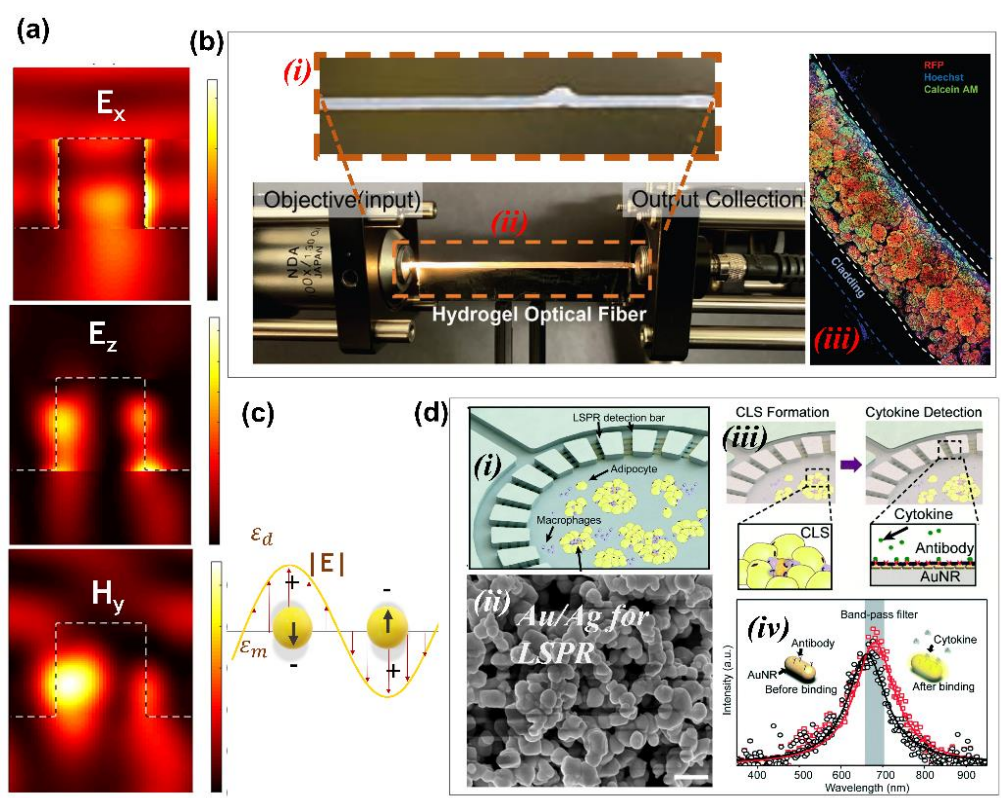

**Figure 3**. Squeezing light. (a) Enhancement of the electric ($E_x$ and $E_z$) and magnetic ($H_y$) fields in the submicrometric waveguides @TM polarization of the incident beam. (b) Fiber coupling: (i) hydrogel fiber, (ii) setup for growing cells and exploiting the waveguide/matter coupling principle; (iii) culturing of the cells inside the fiber. Copyright © 2021 Wiley - VCH GmbH [46]. (c) Localized surface plasmon resonance. (d) Localized surface plasmon resonance-based sensor. (i) Adipocyte culture chamber, which is surrounded by multiple microchannels connected to localized surface plasmon resonance cytokine detection barcode arrays; (ii) an example of nanometric Ag/Au hybrid features that can be employed for localized surface plasmon resonance. Scale bar: 10 nm; (iii) schematic of the on-chip measurements; (iv) binding between the targeted antigen and sensing surface induces spectral redshift and intrinsic intensity enhancement of the scattering light. Panels (i), (iii-iv) Copyright © 2108 Royal Society of Chemistry [54].

The power of a similar system resides in the ability to monitor adipose tissue initiation, differentiation, and maturation and to simulate the characteristic formation of crown-like structures during pro-inflammatory stimulation (Figure 3d(iii)), while allowing multiplex analysis of pro-inflammatory e.g. IL-6 and anti-inflammatory, e.g. IL-10 and IL-4 cytokines secreted by the adipocytes and macrophages (Figure 3d(iv)). A multiparametric SPR biosensor functionalized with artificial cell membranes was used to mimic the environment and perform a two-dimensional affinity analysis of the functional avidity of CD8+ T cells, with a detection limit for flow-through cells about two-fold achieved under the current state of art for plasmonic sensors, a very promising result for the development of personalized immunotherapies against cancer [56]. The proven detection limit for inflow cell detection results has been improved approximately 2-fold over the current state of the art for plasmonic detection. The plasmonic photothermal effect and localized surface plasmon resonance sensing transduction were combined to demonstrate exceptional sensitivity. The probe beam was generated by a broad-spectrum light-emitting diode source operated in the attenuated total internal reflection mode to provide a local plasmonic response. For evidence of the sensing performance, SARS-CoV-2 was studied over the concentration range of 0.01 pM to 50 µM, and it resulted in a limit of detection at about 0.22 ± 0.08 pM and a phase change caused by a local variation of localized surface plasmon resonance confined in a narrow wavelength region [57]. Extensively used for biosensing and relying on localized surface plasmon resonance, the metasurfaces are formed by the assembly of organized nanometric features created with an advantageous nanometric architecture. Surfaces with chiral properties are a specific type of metasurface, and since chirality is a crucial aspect of the molecular domain, these highly functional surfaces play a crucial role in biosensing; DNA, proteins, and numerous synthetic pharmaceutical compounds used in therapy, all contain chirality. The integration of a functional core-shell nano-architecture in a fluid chip enabled the achievement of picomolar sensitivity, which is relevant for the study of neurodegenerative diseases [58]. The proposed configuration exhibits significant flexibility by using a conformal coating with a polymer shell that modifies the near- and far-field optical response of the chiral metamaterial as a result of energy transfer between the polarization charges of the dielectric shell and a free electron of the plasmonic core. The limitations stem from the expensive and time-consuming fabrication techniques required to ensure the periodicity of the structure over a significant active area. Both features are required to stimulate plasmonic Fano resonance transitions that enable multimodal and multiplex detection of tiny biotargets (such as proteins and viruses or extracellular vesicles) [59].

### 3.4 *SPR imaging adapted to microsystems*

SPR imaging aims to provide higher resolution than phase contrast microscopy and interference contrast microscopy. In addition, it is in principle a high-throughput optical detection method that enables the parallel imaging of potentially several tens of channels. A monochromatic or narrow-pass filtered light, passing through a prism typically configured in the Kretschmann configuration, impinges on the activated thin metal surface near the surface plasmon resonance angle. As with prism coupling, activation of the bond between the analytes and the metal induces a change in the refractive index in an environment close to the metal. The latter mechanism provokes an energy loss through the propagation of surface plasmon polaritons, which is then translated into a change in the intensity of the reflected light. Most importantly, the local variation in the refractive index depends on the analyte, and the captured picture obtained through the analysis is related to the temporal and spatial monitoring of surface binding events (kinetic adsorption and desorption measurements). Therefore, the method enables the monitoring of spatial variations in the reflectance of incident light caused by interactions with the analyte via prism coupling and caused by the difference in refractive index. Some examples of combined applications of SPR imaging and grating or localized surface plasmon resonance have been reported above and in fact, metallic nanostructures with a well-designed arrangement integrated into parallel microfluidic channels have reached the target analysis [60]. Triggering SPR imaging via grating microstructures enabled the realization of a miniaturized biosensor platform that is fully integrated into a few-centimeter shell and has a design that can potentially be coupled with the cameras in handheld electronic devices such as smartphones, making it practical to use outside of the clinical laboratory. The demonstrated sensitivity was ~600 RIU$^{-1}$ and was tested against the detection of uropathogenic *E. coli* for bacterial suspensions in buffer solution and human urine, for concentrations ranging from $10^3$ to $10^9$ CFU mL$^{-1}$ with an operation that was completed in 35 minutes [61]. Besides, imaging of SPR was used to detect and monitor antibody-antigen recognition and binding activity using a subnanomolar sensitivity device. The keys to this strategy are multiplexing and high throughput [62], as label-free methods. Microsystems including arrays as well as multilayer structures have been exploited to detect the molecular weight of large as well as small molecules especially proteins (e.g. mouse KIAA proteins (MW ~130 kDa), bovine serum albumin (MW ~69 kDa)) and the sensitivity of the systems of ~1 nM. The flexibility derived from the integration of multichannel with an elevated flow control offer a solid platform where to analyze different analytes, mix them as well as run the protocol for immunochemistry. The application of integrated microsystems including parallel channels and a network of microvalves and micropumps has gained considerable attention for the implementation of SPR imaging and microfluidics-assisted SPR imaging biosensing seems to represent a perfect layout to develop probes designed for high-throughput and multiplex applications [63].

### 3.5 *Plasmonic microsystems integrating arrays*

In the 17th century, Huygens proposed a model in which an array of points scatters the incoming plane wave into circular waves whose phase is given by the envelope of the phase front; this realization successively brought significant development in the field of plasmonics and, from this concept results in the phased array, an arrangement of radiating elements that are phase-shifted with respect to one another. In this manner, the light can be manipulated by leaving the scattering amplitude unaltered but changing the phase of the antenna-like features. Consequently, the scattered circle waves coming out of the antennae merge into a plain wave phase front that is refracted because of the phase difference between the radiations (Figure 4a (i)). Because of the phase shift and gradient in the phase along the surface, regardless of the media's refractive index, the adjustment of the relative phase of these scatters can bend light, as shown in Figure 4a(ii). This allows for a femtomolar limit of detection [64], making phased array antennae applicable for a variety of potential uses that need a signal boost, such as biosensing. A different optical configuration of antenna arrays induces the coupling of the oscillation and triggers a plasmon coupling [65] (Figure 4a (iii)). In this case, the antennae direct the light out of the plane and the resonant scattering cross-section derives from the large polarizability of the antennae confined in a small volume. As a consequence, due to the spatial arrangement of the scattering particles, the emitters' radiation is highly directional and all nearby antenna particles experience plasmon resonances as a result of the single driving emitter. These resonances powerfully drive one another; in fact, the interference caused by any generated currents in the antenna particles forms a highly directional source controlled by geometry [66] and when two metal nanoparticles interact at a distance r, the electric field felt by each particle indicates that the two nearby nanoparticles couple using the near-field equation $E = I_0 \frac{e^{-ikr}}{r} \cdot \frac{\cos[(\pi cos\theta/2)]}{\sin\theta}$. The latter mechanism is polarization-dependent and related to the wavelength, since there can be a shift in the wavelength of the absorption spectra. Arranged in ordered or disordered arrays, the nanometer features produce complex geometries that affect plasmonic behavior with an amplification corresponding to the collective modes and the Fano resonance, which helps tune surface plasmon polariton behavior [67]. Keeping in mind the concept of arrays given above, both the optical arrangement found intersting applications in biosensing and were identically exploited in microsystems.

Phased arrays of nanopores and nanoholes were designed based on optoplasmonic principles of phase matching; their integration into microfluidics has resulted in label-free and multiplexed detection of proteins, carcinoembryonic antigen, fetoprotein antigen, and human epidermal growth factor receptor-2 [68]. Indeed, the nanopores return rapid results, with a limit of detection reaching concentrations as low as attomolar in assay buffer and femtomolar in undiluted untreated human serum. A higher level of sensitivity can be achieved if the nanohole array is resonant at a specific wavelength and designed to match and detune the emission peak wavelength of a vertical-cavity surface-emitting laser before and after binding the molecules. An example is shown in Figure 4b where a plasmonic microfluidic chip (left of the panel) integrates a hole array whose scheme is shown in Figure 4c. The cytokine released from the cells binds the specific antibody immobilized on the surface of the plasmonic sensor and induces a variation in the signal. The plasmon resonance gives rise to a red-shifted signal that depends on the distance of the holes, and the refractive index and it is representative of the concentration of the molecule, which can be measured by monitoring the light output intensity [69]. The integration of nanostructured arrays in a microfluidic environment has been applied to exploit the enhancement brought about by the plasmon coupling and to detect biological samples. In fact, due to more effective near-field concentration and confinement, plasmonic arrays usually show higher biomolecule sensitivity compared to the performance of metasurface dielectric counterparts [70]. They have been applied to both labeled and label-free biosensing to enhance the interaction between light and biomolecules through the effects of near-field confinement. Multiple shapes and dimensions have been integrated into a microfluidic chip to detect the activity of biomolecules. Nanoellipsoid arrays have been used for studying the binding activity of the exosome with a detectability down to 1 ng mL$^{-1}$ [71]. The enhancement of plasmon coupling has been successfully demonstrated for gold and densely packed nanorods organized in a vertically aligned array.

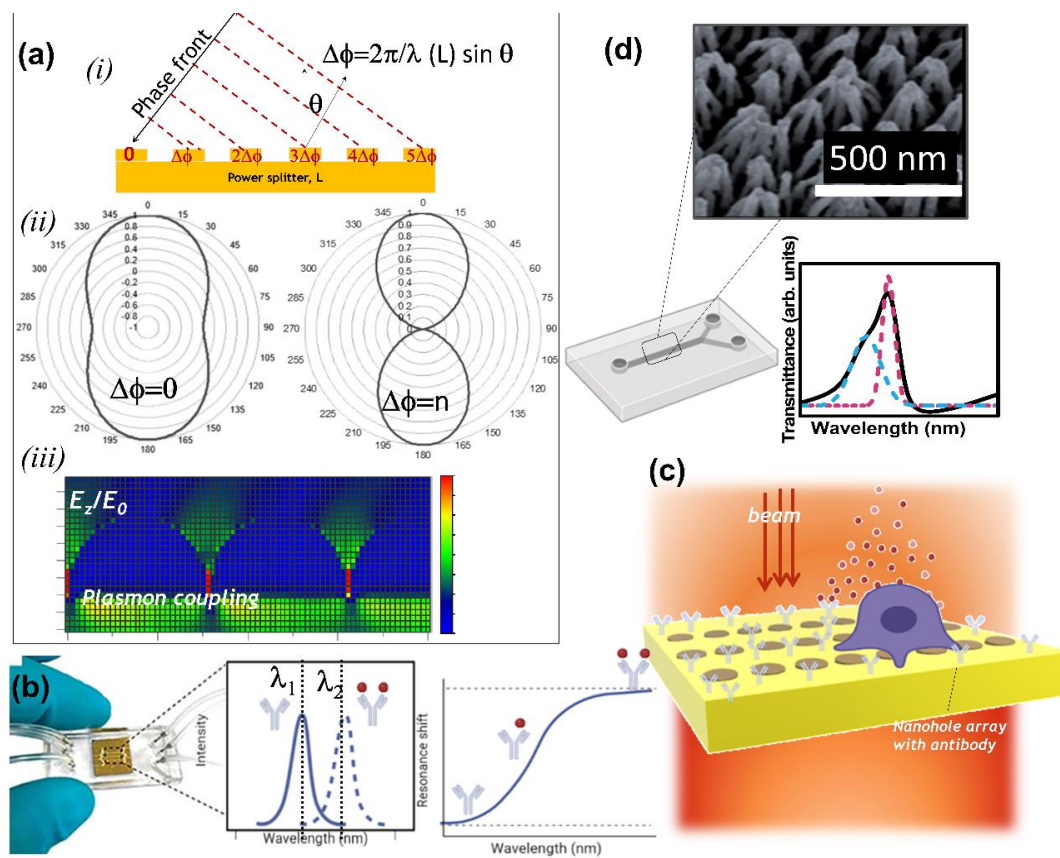

**Figure 4**. Plasmonic array. (a) (i) Schematic of an array of nanometric features respecting the Huygens principle; (ii) Polar plot of emission directivity, for different conditions of phasing, generated by the plain surface $\Delta\varphi = 0$ and in presence of an array $\Delta\varphi = n$; (iii) Enhancement of electromagnetic field triggered by the plasmon coupling. (b) From left to right: Microfluidic device integrating a microhole array for plasmonic sensing. Copyright © 2010, American Chemical Society [72]. A wavelength shifting of the signal before and after the receptor/ligand formation and the transduction of the signal in a sensorgram relative to the binding steps. (c) A possible application of the microfluidic hole array for detecting an anomalous release of cytokines via binding with the antibody immobilized on the surface of the sensor. (d) Top: scanning electron microscope images of vertically aligned nanoantennae assembled in an array and exploited for detecting DNA; bottom left: a microfluidic design that can incorporate the features; bottom right: transmission signal of the system. Copyright © 2018, American Chemical Society [73].

The optomechanical configuration not only acts as an optical antenna by receiving, limiting and amplifying the external signal but also provides an excellent environment to study the binding kinetics of different DNA strands as well as their hybridization with the ability to signal transduction of the hybridization of two different and complementary strands. The latter investigation was developed by exploiting the nanofeatures shown in the microscopic image of Figure 4d [73], and integrated into a microfluidic chip (shown schematically in Figure 4d). The results of this application demonstrated that upon immobilization of the strands on the surface of the microfluidic layer, a 1 nm red shift was detected, while an additional 2 and 5 nm of backfilling and hybridization resulted in a peak shift of 25 and 100 paired bases; a schematic diagram of the displacement is shown in Figure 4d (bottom panel).

### 3.6 Plasmonic microsystems with electron density tuning

All of the observations summarized above have led to the design of very complex metal nanostructures that can lead to the control of light properties, with consequent advances in related application areas. However, one can argue that the flexibility of these systems is modest as it is closely related to the design and manufacturing facilities, while the large ohmic losses remain unresolved. This observation has prompted discussion about the need to provide ample flexibility to the plasmonic structures, and it has brought to the tuning of the electron density and related extinction spectrum developing a method that involves manipulating material properties rather than light parameters. One of the most important properties of metals is that they have an extremely high carrier density, namely one free electron per atom. The change in carrier density that can be achieved is only a few unit percent, so there is not much latitude to adjust the absorbance spectrum. In contrast, since the semiconductors do not have a negative dielectric constant and have a lower electron density, there is a limited demand for them as plasmonic materials. Intermetallic materials such as conducting oxides (e.g. graphene [74,75], indium tin oxide ITO [76]) have an intermediate value of electron density. Their intrinsic properties make them an excellent model for plasmonic applications since the real part of the permittivity can show a transition from positive to negative values, as it moves from the infrared to the visible region. With the control of tunability by chemical and electric doping, the charge density varies and causes the electromagnetic properties of the material to change from metallic to dielectric [77].

### 3.6.1 Chemical doping for enhancing plasmonic biosensing

Among the materials that can be chemically hybridized, graphene is the most interesting because of its chemical structure. Mechanisms of graphene doping are shown schematically in Figure 5a. The donation of an electron (n-type doping) or a vacancy (p-type doping) from a dopant modifies the electronic density of the graphene; consequently, the energy gap raises with a shift of the Dirac point. This electronic modification affects plasmonic behavior. Graphene and doped graphene are very attractive because of the typical monolayer structure that allows a simple integration in lab-on-a-chip and microfluidic systems [78–81]. Furthermore, using graphene or graphene oxide as a functional layer between the metal film and a biological layer has a strong potential to improve the sensitivity of biological measurement by turning on and off the mechanism of doping. The added value of integrating graphene and graphene-based materials into microfluidic devices is to combine the high sensitivity of the material with the low cost of the microfluidic devices, allowing easy access to fast and accurate tracking. Moreover, since graphene and graphene oxide are hybridized with metals, the hybrid architecture usually shows two split modes that are correlated with the sensitivity of the systems [82]. Plasmonic and lossless behavior has been shown by the stratification of a high refractive index–graphene oxide and polymethyl methacrylate, the latter having a low refractive index. Peculiar behavior was recorded relative to the aggregation of amyloid-beta (Aβ) protein into amyloid fibrils, considered a major hallmark of Alzheimer's disease. The open fluidic system shown in Figure 5b (left) was used to identify Aβ16–22 aggregation. The mechanism of aggregation of the protein causes the graphene doping by the reduction of π-π stacking between graphene and Aβ16–22 peptides; this induces an increased Rabi splitting between the 2D and G peaks of the graphene. It was observed that Rabi splitting of the spectra peaks increased due to the graphene electron doping and Fermi energy level variation corresponding at each different phase of Aβ16–22 peptides aggregation (Figure 5b, right), providing a method of the aggregation monitoring tracking the position of Rabi splitting peaks [83]. Plasmon-coupled emission from graphene oxide reveals the existence of strong surface states on the top layer of the photonic crystal framework. The chemical defects in the graphene oxide thin film trigger plasmon hybridization within and across the multi-stack leading to an elevated sensitivity of the platform and a limit of detection measured at 1.95 pg mL$^{-1}$ for human IFN-γ [84]. A special case of electron density tuning is represented by the liquid metals [85], which have started to attract great attention in the last decade, with interesting implications in plasmonics, as well as in the design of a dynamic environment in microfluidics (e.g., droplets). In this context, a 3D plasmonic oxide framework was realized by eutectic gallium-indium liquid metal droplets with an ultrathin porous hexagonal molybdenum oxide in an angstrom scale. A large number of oxygen vacancies of the hexagonal molybdenum oxide leads to a charge concentration capable of inducing a broad surface plasmon resonance over the entire visible light spectrum, with a promising positive effect on the spectroscopic behavior with an amplification factor of up to $6.14 \times 10^6$ [86].

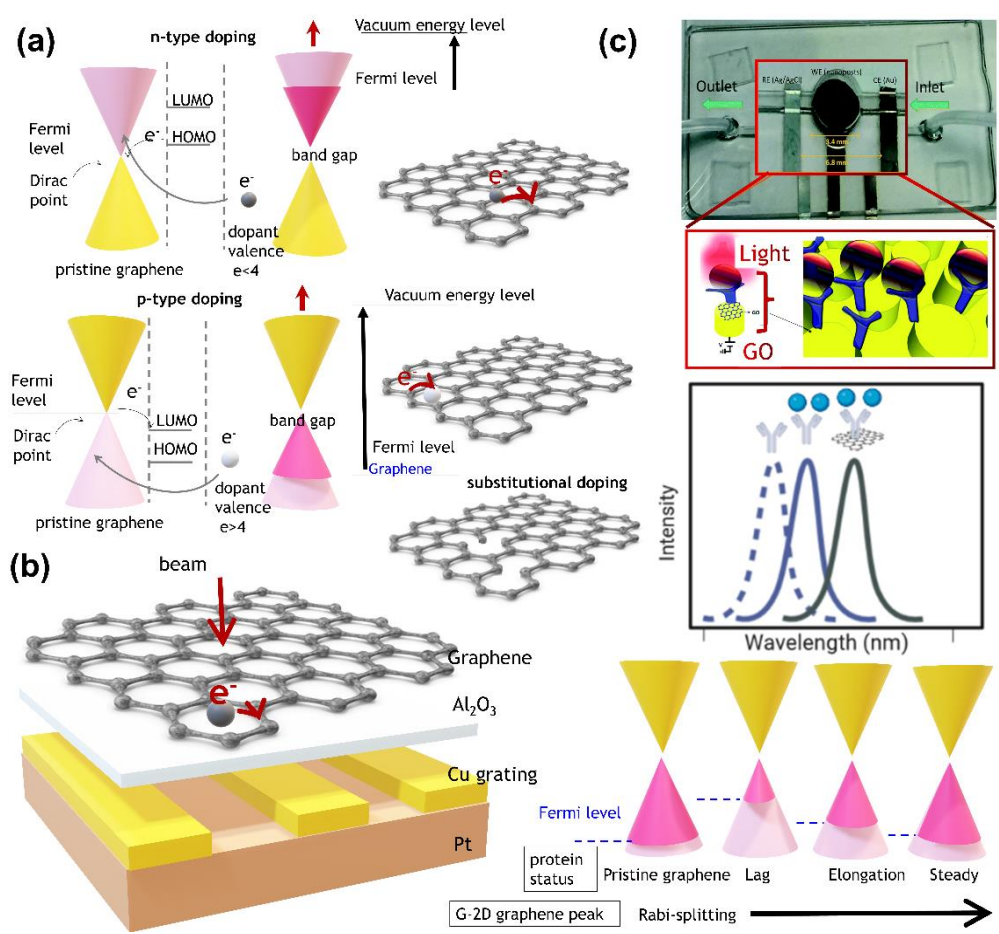

**Figure 5.** Graphene-based plasmonic biosensors based on electronic density tuning. (a) Mechanisms of graphene doping relative to the n-type (top) and p-type (bottom) doping with a relative donation of an electronic vacancy or an electron from the dopant, respectively. (b) Left: An open fluidic system based on graphene. Right: Fermi levels are illustrated before and after the chemical doping by the protein doping. The arrow indicates the direction of increased Rabi splitting. (c) From top to bottom: dual-modality sensor chip. Inset: zoom on the sensing area, schematic of the working principle, and the wavelength shift of the plasmonic signal after the antibody/ligand binding (inset in the bottom figure) for the pristine and graphene-based sensor. Copyright © 2108 Royal Society of Chemistry [87].

### *3.6.2 Electrical doping for enhancing plasmonic biosensing*

The electrical doping of graphene is one of the most promising and exploited strategies for realizing fully integrated sensor systems; being compatible with microfluidic and lab-on-a-chip technologies, it has shown fairly interesting results. The general scheme consists of a transistor whose gate is made of graphene or tin indium oxide. A noble metal forms the electrodes, so a negative/positive gate voltage shifts the Fermi level relative to the Dirac point, according to the scheme in Figure 5a. Under this condition, it can be seen that by selecting the gate carrier density in the conducting oxide, the application of the voltage tunes the permittivity of the gate material from positive to negative before destroying the insulating gate and offers a wide degree of flexibility for these systems. The mechanism triggers significant implications for the tunability of the optical properties via the application of the gate voltage [69]. The graphene field-effect transistor is usually monitored through the Dirac voltage shift, which has also demonstrated variability with the fluidic flow at the sensing interface. To increase sensitivity, dual tracking was done generating both electrochemical and surface plasmon resonance signals from a single sensing area of Au/graphene oxide nanoposts (Figure 5c, top) [87]. The functionalized Au/graphene oxide nanoposts fabricated by soft lithography function as a spatially well-defined nanostructured working electrode for electrochemical sensing, as well as a nanostructured plasmonic crystal for SPR-based sensing through the excitation of surface plasmon polaritons (Figure 5c, middle). Plasmonic detection has enabled the tracking of antigen-antibody dynamic interactions as well as the extrapolation of the association and dissociation phase, mechanisms that occur at the sensor surface. When compared with a pristine sensor (e.g. naïve Au), the Au/graphene oxide sensor displayed an enhanced shift (Figure 5c bottom). On the other hand, fluidic control offers an additional degree of freedom that can be tuned to optimize the sensitivity of the transducer. For example, exploiting the fluid dynamics of a flow cell design including an impinging jet geometry has offered high sensitivity for the detection of DNA hybridization (~44 mV/decade of target DNA concentration) and a detection limit of ~0.642 aM [88]. The conjugation of the graphene field-effect transistor with anti-CD63 antibodies for the label-free detection of exosomes has brought to an additional minimum alongside the original Dirac point in the drain-source current- back-gate voltage curve. The phenomenon was observed due to the change in the electrical properties of the exposed graphene. The sensitivity of such a system, equal to 0.1 µg mL$^{-1}$, was associated with the shift of time when the exosomes come in contact with the antibodies at the graphene surface [89]. Analogous experiments were carried out to detect lysozyme by evaluating the concentration-dependent calibration curve and the dissociation constant, $K_D$=375.8 pM. Upon varying the concentration of the protein, the sensitivity of the system was estimated at 2.6 pM [90]. Systems based on graphene field effect transistors have also been used to detect SARS-CoV-2 with a limit of detection of 1 fg mL$^{-1}$ after the spike antibody was applied to the graphene [91]. The main strength of those systems derives from the extremely high sensitivity that allows the detection of the virus in the transport medium used for nasopharyngeal swabs from both cultured SARS-CoV-2 samples as well as from SARS-CoV-2 clinical samples [92].

### 3.7. Quantum plasmonic sensing and microsystems

#### 3.7.1 General concept of quantum plasmonic sensing

The final section of this review introduces quantum plasmonic biosensing, a promising approach that has the advantage of requiring less optical power, resulting in less chance of damaging the biological samples under study [93,94]. In recent years, quantum technology for encoding quantum information has reported evolution across several frameworks and has been classified according to the degree of interaction with the environment, which was intended as a measure of how the system controls and protects quantum information [95]. The role of photons is essential because they represent the scaffolding that better protects the quantum information during its propagation, either in vacuum or in matter [96–98]; besides, quantum photonic processors allow the creation and manipulation of the qubits to encode the photons for universal two-qubit quantum computation [99]. The downside is that photons slightly interact with each other and with the surrounding environment, resulting in inevitable losses of the photons and a quantum communication rate that falls off exponentially with distance.

#### 3.7.2 Light/matter coupling for quantum plasmonic biosensing

Despite this inherent problem of the photons, it is still possible to construct quantum networks at high speeds by enhancing the interaction at the nanoscale between light and matter, of which the degree of mode coupling is correlated to the sensitivity of plasmonic biosensors. Then, due to the confinement and amplification of light within a cavity or nanogap structure, a strong coupling has the potential to produce higher sensitivity with less optical power [100]. Because of their mechanism of generation (Figure 6a), plasmon polaritons rely on the coupling between light and matter and have a quantum character that develops in energy losses related to a mechanism of hot carrier generation, which affects the measurement of the molecules attached to the surface of the quantum plasmonic sensors [101]. The interaction of light and matter intrinsic in plasmon polaritons is the key concept of the quantum electrodynamic cavity. The fingerprint of the efficiency of light/matter coupling finds in several factors the characteristics of its strength. The mode volume $V/V_0$, which accounts for the spatial distribution of electromagnetic energy surrounding the matter, and the quality factor $Q$, which measures how long photons and emitter interact, are correlated to the photon generation rate that accounts for the decay rate of the radiative excited state. The narrowing of the spectra width indicates a situation where the generation rate is faster than the decay one and corresponds to a strong light/matter coupling. Arranging the atoms and emitters in a resonator-like cavity enables the coupling between the resonator mode and the emission modes; in this way, the spectral width of the emitters grows and overcomes the cavity that in turn displays spectral width shrinking and narrow emission linewidth, which influence the directionality of the radiation. Figure 6b schematizes the main regimes of light/matter interaction [102]; the left side of the top panel displays an optical resonator with a single two-level atom, namely a qubit (left), while the right side shows the same optical resonator coupled to a number N of quantum emitters. The strength of the coupling is estimated through parameters such as the resonance frequency of the cavity mode $\omega_c$, the transition frequency of the qubit $\omega_q$, the coupling strength $g$ and the loss rate of the cavity and qubit $\kappa$ and $\gamma$; while for the system including N emitters, the strength of coupling depends on the coupling with the single qubit and all emitters, $g\sqrt{N}$. The bottom panel of the figure displays the light/matter coupling regimes, from a weak condition derived by the tuning of the electromagnetic environment according to a Purcell effect up to a deep strong coupling generated by the effect of THz metamaterial coupled to a cyclotron resonance of a 2D electron gas [103]. The quantification of the coupling strenght has reason to be defined, because regime of strong and deep strong coupling results ubiquitous in quantum phenomena and their quantification. Raman spectroscopy represents an example of this concept, then the next section will display some recent findings.

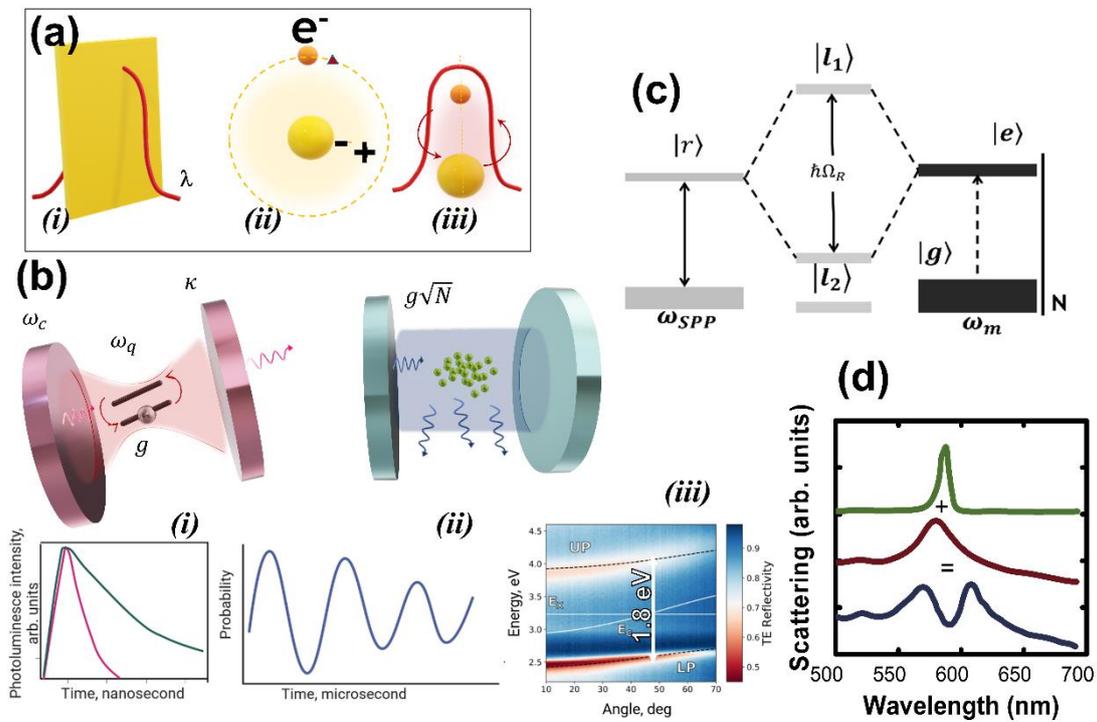

**Figure 6.** Light/matter coupling. (a) Plasmon polaritons as a naïve model of light/matter coupling. (i) When the optical wave wavelength $\lambda$ passes through the metal releases energy to the electron that starts to rotate around the atomic nucleus that becomes a hole (+) (ii).The hole and the electron have different signals, and during the time the electron rotates around the hole they form the exciton. The exciton recombines when the electron falls back into the hole and the energy is given back to the optical wave, with the same frequency as the one that created the exciton; the conversion exciton/wave cyclically continues. When the metal is under resonance, that is the metal is in the middle of the single wave, the exchange light/exciton creates the polaritons (iii). (b) Top: an optical resonantor with a single two-level atom, qubit (left), and the same optical resonantor coupled to a number N of quantum emitters (right). Bottom: (i) Weak coupling displayed by the dynamics of the spontaneous emission of a qubit (e.g. quantum dot) by a cavity. Photoluminescence intensity relative to a qubit on resonance (faster qubit decay) and off resonance (slower qubit decay) with the cavity displays Purcell effect. (ii) Strong coupling resulting in vacuum Rabi oscillations and deriving from Rydberg atoms coupled to a superconducting microwave. The atom in an excited state enters an empty resonant cavity and oscillate before decaying. (iii) Deep strong coupling: experimental angle-resolved, TE-polarized reflectivity map for a cavity containing aligned oriented poly(9,9-dioctylfluorene) Copyright © 2020, American

Chemical Society [103]. (c) A hybrid model for coupling SPP/n molecules. (i) The excitation state of the molecule is labeled as $|e\rangle$ and ground state $|g\rangle$ strongly couples with plasmonic resonance state $|r\rangle$, and causes the levels repelled by a Rabi splitting. (d) Scattering spectra of the strong coupling showing the Rabi splitting (bottom curve).

### *3.7.3 Strong light/matter coupling for surface enhanced Raman scattering*

Surface-enhanced Raman scattering SERS and Raman-based spectroscopy require their chapter in this context. According to the classical interpretation, Raman phenomena are directly related to the localized surface plasmons that amplify the near-field signal, and they are correlated to an electromagnetic enhancement produced by a nanometric feature [104] [105] or mesoporous nanostructures [106–109]. Therefore, nanometric particles have been integrated into microfluidic channels in real-time and in a continuous flow, enabling the discrimination of several analytes, biomolecules, as well as viruses. There is a plethora of applications relative to Raman spectroscopy exploiting the characteristics of different microsystem designs. *L. monocytogenes* and *Listeria innocua* in just 100 seconds were identified by using Au nanostars in a microfluidic environment [110], and Au nanoparticles were used for estimating β-2-microglobulin by the binding with the anti-β-2-microglobulin attached to the particles [111]. The latter shows a specific coloration due to the localized surface plasmon resonance of the nanoparticles when filled inside a microfluidic cuvette designed for efficient mixing of the analyte loaded with β-2-microglobulin. The noble metal has been combined with $Fe_3O_4$ [112] or graphene oxide [113], while microfluidics has been fully exploited for developing a microdroplet assay for single-cell analysis. In this last application, in particular, Ag and magnetic beads were exploited for turning on the SERS signal by the cytokines secreted by the single cell (limit of detection of 1.0 fg mL$^{-1}$ in one droplet) [114]. Fluidic SERS has been used to locate the biomarker of important clinical disorders like infarction and cancer in fully integrated devices [115,116] [117] [118].

A review of the state-of-the-art in Raman spectroscopy in microsystems reveals two areas that appear to be subject to criticism: (1) the transition from academic to commercial use; and (2) the requirement to interpret the method in terms of strong light/matter coupling. Regarding point (1), it's interesting to note that the majority of the suggested protocols are still in the research phase despite the many exciting results. Due to a lack of trustworthy technologies and well-established methods, the technology is only used in research labs. Currently, a number of issues prevent practical application on an industrial scale, including high manufacturing costs brought on by processes with poor scalability, poor batch-to-batch reproducibility, substrate stability, and homogeneity. Although SERS/microfluidics have demonstrated excellent results, and both SERS spectroscopy and microfluidics are growing in popularity, there is significant interest in adopting these technologies in the industry. The development of dependable, affordable disposable SERS microfluidic chips employing high-throughput manufacturing techniques like injection molding and roll-to-roll [119] as well as the use of transparent polymers for the construction of microsystems [120] are examples of efforts in this approach.

The second point listed above is related to the light/matter interaction strength. It has been demonstrated above that the limitation of the photons' inability to interact naturally has been highlighted above; this limit appears to be overcome by an intermediary light/matter interaction, on which the notion of quantum electrodynamics has been developed. The strong light/matter coupling of plasmonics has proven to be sensitive enough to identify single molecules, indeed molecules pose a stronger coupling challenge than those containing quantum emitters due to their Stokes-shifted emission and a greater variety of vibrational modes [121–123]. The sensitivity depends on the excitation electric field and the emitter dipole momentum, both of which are related to the strong coupling with the molecules that leads to the so-called Rabi splitting (Figure 6c and Figure 6d). The sensitivity of systems with strong coupling can be improved by several orders of magnitude through optimization [124], making it possible to analyze extremely low analyte concentrations [125,126]. An interesting and recent explanation of Raman phenomena places the spectroscopic technique among the more sensitive and motivates this theory by explaining Raman in terms of a mechanism of strong light/matter coupling [127]. It has been observed that the inelastic interaction of molecules with optical photons is the fundamental concept that Raman scattering uses for detecting molecular vibrations, but the low efficiency of the readout necessitates a significant number of molecules. By a mechanism of interaction between the plasmonic and the vibrational modes, surface-enhanced Raman scattering displays a sensitivity that is enhanced by several orders of magnitude. The coupling between a single molecular vibration and a plasmonic mode represents an optomechanical interaction, analogous to that existing between cavity photons and mechanical vibrations and it brings the optomechanics down to the scale of molecular vibrations. The molecular optomechanical description of SERS is new and although there are several interesting theoretical approaches, applications are still lacking. Among the promising advantages brought by a quantum description of SERS there is the capability to detect and analyze the statistics of the emitted photons that allows for the interpretation of two-color correlations of the emerging photons and the generation of nonclassical states of light. The reinterpretation of SERS as being based on a strong light/matter coupling regime confers to the method an extraordinary molecular sensitivity down to a zemptomolar range and can reveal a variety of Raman transitions that are typically invisible with the classical interpretation of Raman. This has made attractive the design of nanofeatures that together with the molecules work as mechanical oscillators and interact with the light [128]. The majority of the research efforts in this area are still concentrated on creating numerical and analytical models to obtain the requisite expertise [129]. Unfortunately, even with the advancement in theoretical modeling and analytical results from embedded quantum emitters and antennae in resonant cavities and the ongoing trend toward on-chip integration and miniaturization to realize optical signal processing and integrated circuits, there are still some significant performance gaps that must be filled before using them for practical applications.

### *3.7.4 Quantum plasmonic microsystem biosensors*

After a conceptual premise related to the concept of light/matter coupling, this section continues describing quantum plasmonic biosensing and highlights the integration in nano and microsystems, taking into consideration that with the contribution of the flourishing micro- and nano- technology the proposed solutions have the potential to become more and more sophisticated. Quantum plasmonic design for metrology have several configurations, but the fundamental metrology architecture consists of a classical detector for the quantum states of light, while the detection principle integrates the coherent photon-plasmon conversion processes or the signal from non-coherent source. With the attention to the cavity and hybrid cavity mode, several solutions have been proposed that span from a multilayer up to cavity/antenna configuration. Therefore, metal-dielectric configurations have been developed to promote strong light/matter interactions down to the single photon and achieve a propagation mode within the diffraction limit [130] [131]. The optical architecture demonstrated an improvement from the integration in fluidic microsystems enabling a 90% increase in photon collection efficiency thanks to which, in the recent applications, the molecular analysis has been carried out. For example, for developing a glucose sensor, a cavity mode has been successfully combined with the one of an optical fiber made of a smart and flexible microgel [101]. The fiber/cavity integration conferred a tunable character to the cavity, because by swelling/shrinking gel mechanisms an intertwined interaction between the plasmonic resonances and cavity modes is turned on and off. It is worth noting at this point that the phase singularity that depends on the topology of the cavity surface and enables great efficiency when attained cannot be achieved by the aqueous system required for the biological samples even with a high efficiency of the system of measurement. The achievement of a single mode is an important challenge for the design of the fluidic microsystems [132] [133]. because its achievement would be followed by and enhancement of the sensitivity down to the single molecule as well as for single cells. An alternative to circumvent the obstacles related to the integration of the optical couplers derives from the selection of materials such as the silicon, which from one side has the advantage to be biocompatible and from the other enables the achievement of a quality factor (e.g. $Q>10^5$) and simultaneously a high step change in resonance. In fact, similarly designed devices have demonstrated a confined cavity mode that allows the monitoring of the binding activity of individual adenoviruses that has a typica diameter of 100 nm. Systems like photonic cavities display a decrease efficiency due to a low radiation rate. To overcome the decrease of the radiation rate of a photonic cavity and improve the efficiency of the measurement, the cavity has been hybridized with optical antennae, which allow an effective electromagnetic field confinement and photon emission faster than the lasting time must be triggered inside the matter. This duality of behavior has been achieved by systems featuring an antenna mode and a cavity mode; the cavity realizes a high efficiency in photon collection by holding the radiation and enhancing the light/matter interaction, while the antenna improves the directionality of the emission, by coupling the system to the free space [134]. The hybridization of the two modes has the ability to push the detection limit down to a THz regime and to induce the matching of two impedances, resulting in a reduction of the lossy character of the metal and linewidth [135], with a significant impact on the radiative behavior as well [136]. As an example, Figure 7a shows a waveguide (panel i) fabricated using a combination of bottom-up/top-down processes; the device is strain tuned and was demonstrated when integrated with a quantum emitter and a planar integrated optical resonator consisting of a silicon nitride ring resonator fabricated on a piezoelectric substrate (panel ii). The tunability of the devices was tested by applying a voltage to the piezoelectric substrate while the quantum dot emission collected through the waveguide was recorded with the focusing optical fiber (panel iii) [137,138]. Ideally, an adiabatic coupling between the plasmon polaritons into a dielectric waveguide enables an efficient collection and simultaneously a significant enhancement of the sensitivity; the two conditions have been implemented in the photonic cavity like a whispering gallery mode that integers plasmonic structures in an opportune architecture for achieving the enhancement [139] [140]. A simplification of the whispering gallery mode in microfluidics is shown in Figure 7b(i) where the whispering mode is generated between a sphere and the Au nanorod the is permanently adsorbed on the sperical surface; an example of expected shift of the transmitted signal deriving by the hybrid mode is shwon in Figure 7b(ii). The SPP modes that lead to the propagation of electromagnetic waves along metal-dielectric or metal-air interfaces transport a big number of high-quality information than the ones related to the photonic waveguides, which only can convert between photons and plasmons. The reduction of the detection limit down to zeptomoles with a correlated increment of sensitivity has been recorded using an original optical configuration including nanohole and nanodisk structures. An excellent platform for high-performance chemical and biological sensing is a nanoring resonator array, which displays narrow-linewidth spectral characteristics with a high peak-to-dip signal ratio and significant near-field electromagnetic amplification. In fact, within a few nanometers, the light generates strong electric fields that amplify the scattering of adsorbed molecules and produce a rich fingerprint spectrum with all available chemical information. Ridge waveguides and grating far-field couplers have been exploited to integrate quantum emitters made of organic-molecule with the aim to combine quantum light sources and nonlinear elements that represents a step toward photon-based quantum information and communication. The cartoon of the device is shown in Figure 7c (i) [141]; it displays that a molecule embedded in anthracene is deposited on the waveguide with a coupling efficiency of up to 42 ± 2 % over both propagation directions. As a result, the analysis of the signal evidences that the intensity grows because of the waveguide/molecule coupling (panel ii) more than for a condition of the molecule off guide. It is important to note that in these complex systems, which did not possess a quantum intrinsic nature, the interaction between light and matter was stimulated and amplified by nonlinear effects and interactions of photons with the surrounding matter as well as by instantaneous electric field enhancement imposed on the photons by brief pulses [132,142]. Alternative techniques for single-photon emission have been demonstrated, based on the coupling of optical fields with quantum emitters like a defect in a solid, atom, ion, or quantum dots. These sources can capture and release a single photon, however the isotropic emission makes photon collecting more difficult and reduces sensing efficiency.

If coherent control over light/matter interactions at the single-photon level was a driving force for quantum plasmonic sensing, coherent with the process being pursued for information technology, nonclassical sources have received more attention than promising sources in the recent past obtained for quantum noise reduction. The quantum states of nonclassical light that are highly favored in quantum photon detection are squeezed states [143] and entangled states [144].

The latter, the photon entanglement, allows a quantum state of photon pairs that cannot be described as single one. The squeezed states of light are attained by the redistribution of the observable uncertainty and in fact the advantage occurs by reducing the uncertainty in the variance of one quadrature, namely the statistical noise of the phase $\Delta\Phi$ or of the amplitude $\Delta I$ of a quantum field is reduced at the expense of an increased noise in the conjugate variable according to this correlation $\Delta\Phi\Delta I = \hbar/2$. It follows an increase of the signal-to-noise ratio when detecting physical phenomena that transduce the squeezed variable. The integration of quantum plasmonic sensing in microsystems aims to provide a solid framework for ensuring sample stability and subsequently limiting all sources of noise and signal contamination [97]. Examples with a parametric single-photon source using microfluidics and quantum sensing to measure protein concentration were reported and it has been demonstrated that devices with a similar optomechanical design offer the advantage of combining the stability of integrated optics with the high-precision handling of liquid samples. Moreover, micro- and nano-systems offer the advantage of reducing the uncertainties and oscillations caused by handling the sample, thereby reducing measurement noise as reported by recent investigations. The optomechanical setup displayed in Figure 7d(i) offers an example of an integrated measurement system that includes an idler and a coincidence point for the comparison with the signal [145]. The variation of the analyte refractive index was determined through the estimation of the statistical error and an esemplification of the trend of the signal to noise ratio versus the refractive index are shown in Figure 7d(ii) (panel 1-3 from left to right). The signal and the idler have the same photon number, but the distribution of the idler and source is Poissonian. The statistical analysis operated on the number of the transmitted single photons for a certain number of times to enable the estimation of the standard deviation and demonstrate that estimated errors beat the signal noise limit that would be obtainable by a coherent state of light with the same average photon number as the single photon. Therefore, like the coherent approach, the quantum approach to plasmonic detection analyzes the variation of the transmitted or reflected signal after propagation through the sensor. However, the quantum benchmark considers the photon number state $|N\rangle$ rather than the photon number N, making the optimal state for single-mode transmission and the magnitude of quantum gain independent of the number of photons, as shown in Figure 7d(ii) (panel 4 right) [94], and the statistical analysis is performed on the number of individual photons transmitted. A similar approach allows the standard deviation to be estimated and, indeed, it was shown that the errors are smaller than the signal noise limit, which in turn would be possible for a coherent light state with a mean photon number equal to that of a single photon. Therefore, in light of these results, the sensible further development suggests combining signal and idler analysis on a single chip for a compact solution.

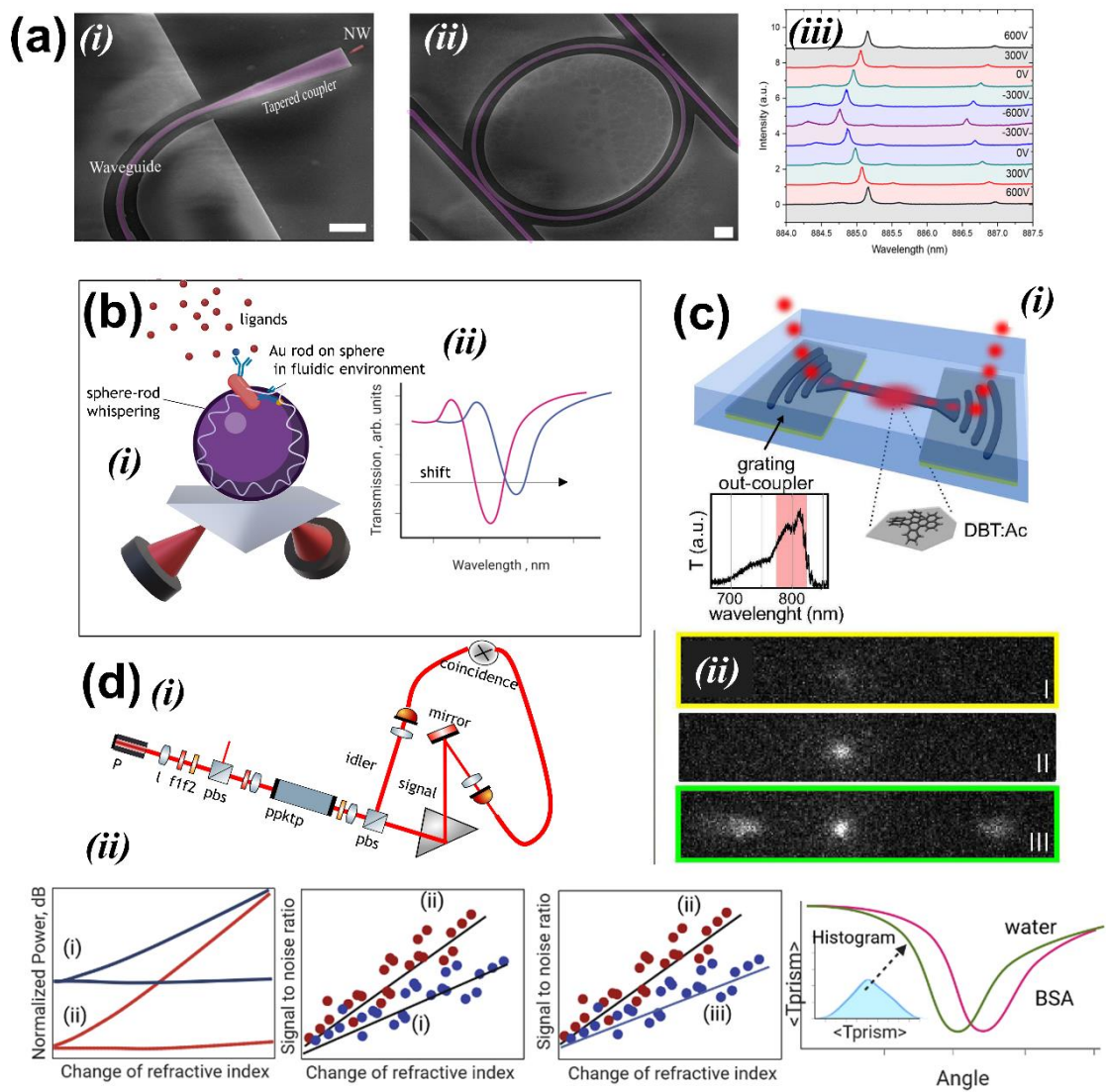

**Figure 7.** Microsystems and quantum sensing. (a) (i) scanning electron microscope image of a silicon nitride waveguide in purple (scale bar 2 μm); (ii) scanning electron microscope image of silicon nitride ring resonator (scale bar 2 μm); (iii) emission spectra of the nanowire quantum dot collected from the waveguide as a function of the applied voltage to the piezoelectric substrate. [137] © 2018 American Chemical Society. (b) Whispering principle: (i) schematic diagram of an optoplasmonic sensor made of sphere and Au rods; (ii) shift enhancement. (c) Waveguide/molecule coupling: (i) cartoon of the waveguide with two grating couplers. A molecule embedded in anthracene is deposited on the waveguide. Inset: grating optical response of the molecule (DBT); (ii) image of the fluorescent signal by excitation with a diffraction limited laser spot. Scheme I: no molecule on waveguide, II: molecule off guide, III coupling waveguide/molecule. [141] © 2017 American Chemical Society. (d) Signal/idler quantum sensing. (i) A single photon is the input in a plasmonic ATR sensor with the Kretschmann configuration, where the actual reflection of a single is treated as transmission through the ATR setup. The generation of the single-photon state (the signal) is heralded by the detection of its twin photon (the idler), by a quantum correlation of photon pairs initially produced via spontaneous parametric down-conversion. (ii) Sensitivity enhancement of plasmonic sensor with quantum resources when probing with coherent states. Left: measured signal while linearly ramping the driving voltage of the ultrasonic transducer that changes the refractive index of air; Middle panels: signal-to-noise ratio when probing the plasmonic sensor versus the change of the refractive index (plasmonic sensor with coherent states trace (i), with twin beams trace (ii). Trace (iii) gives the estimated SNR for the single coherent-state configuration). Plots adapted from ref. [145]; Right panel: Measured transmittances $\langle T_{prism} \rangle$ over the incident angles for BSA and high purity water. The error of the plot is measured as a standard deviation of $T_{prism}$ in the histogram embedded in figure. Plot adapted from ref. [94]

## 4. Conclusions, outlook and future aspects

The plasmonic technology is based on the principle of nanooptics and is theoretically well-compatible with the concept of miniaturization and point-of-care diagnostics. Indeed, among others, plasmonic biosensing, which naturally relies on optical phenomena limited to the subwavelength scale, has found in microsystem integration an optimal environment in which to trigger the excitation of the mechanism of sense. On the other hand, micro- and nanotechnologies have advanced to an essential point in their development in recent years, allowing for the fabrication of devices with a size of a few hundred nanometers that are intended for sensor integration as well as enabling the rapid development of bioanalytical methods. The flexibility of the microsystems environment has become a favorable environment to employ advanced analytical technologies and study biomolecular behavior and cellomics. Since the first example of integrating the plasmonic sensor into a microsystem, gratings, waveguides, nanometric particles, and features as well as arrays have been incorporated into more or less complex and compact designs. Some plasmon excitation strategies (e.g., gratings, waveguides, arrays, and metasurfaces) have found the greatest merit due to their better compatibility with the microfluidic environment; in turn, some optomechanical configurations require more effort to align the transducers and the source. This combination has brought several advantages, some of which strictly depend on the intrinsic nature of the miniaturized environment, others stem from the plasmonic/microfluidic overlap, such as real-time and label-free monitoring of molecular binding events, with interesting advantages in terms of cost reduction, high sensitivity and specificity, fast response and the possibility of multiplex detection.

Although there are now certain restrictions on this development, plasmonic sensors have the potential to become a point-of-care diagnostic tool and account for a sizable portion of the label-free sensor industry. The device must compete with other exceptional technologies and be compact, accurate, reliable, repeatable, cheap, and disposable in order for the technology to be translated on a wide scale. These accomplishments are not insignificant, but they are definitely doable. From a critical stance, two primary issues—the sensitivity of the miniaturized integrated systems and the costs—explain why plasmonic sensors for point-of-care have not yet attained a high degree of industrialization. The latter is the process of getting the product to market; it involves a variety of characters and procedures that determine whether the product will be successful on the market. There are several reasons why plasmonic point-of-care sensors have not yet found a lucrative market. Some of the proposed plasmonic devices are still plagued by signal interference that reduces the binding signal's specificity. This is in contrast to label-based approaches' great specificity (e.g. immunoassays). Furthermore, the current state of plasmonic instruments is not suitable for production aimed at the mass market as they are still cumbersome and expensive. However, there are some exceptions provided by instruments produced by companies such as Biacore, Texas Instruments and Seattle Sensors, although the chip available for those instruments is still expensive and uncompetitive with the disposable devices used for diagnostics.

Since the chip continues to be the primary expense for plasmonic sensing, the process of translating research findings into practical applications must take great care to consider the device cost. Most plasmonic sensors such as arrays and nanoholes require expensive lithographic applications that have notoriously low throughput and high cost and then increase the cost of the chip while hampering transfer from laboratory to clinical application.

The effort to support the spreading of plasmonic measurement in clinical applications is motivated by the number of advantages that the plasmonic sensors show. These include the ability to detect the equilibrium constants of the kinetic rates of biomolecular interactions, which are crucial for fundamental molecular analysis; albeit, even more important are the advantages of direct detection and the speed of plasmonic sensors. Advantages such as flexibility and rapid testing, therefore, make this class of sensors extremely well-suitable for applications related to large-scale and widespread deployment. These aspects do not seem to be the decisive factor that allows the plasmonic sensors to replace other bioanalytical methods like the dipstick tests, but they can offer support in the market niche that is not yet filled with a valuable solution. I assume that the exploitation of these advantages will entail the development of two development paths of plasmonic point-of-care sensors: (1) personalized medicine, (2) critical environment (underdeveloped countries, emergency states, etc.). In the first case, future point-of-care plasmonic testing methods that are reliable and easy to use will contribute to the advancement of tailored medicine; the creation of trusted ambulatory technologies for medical diagnosis of infectious diseases would have significant implications for disease surveillance and treatment (e.g. wearable plasmonic band ). On the other hand, viral infections and the detection of toxins in water and food under difficult environmental conditions are becoming target applications for sensors that do not require long and complicated sample preparation protocols. The development perspective of the plasmonic sensors must have several goals, such as reducing the production cost of the batch devices and increasing the reproducibility, but also aim to become a fully integrated and wearable sensor platform to stand out from the needs discussed previously. For fundamental studies, a promising result comes from the application of quantum plasmonics, which has shown lower optical power consumption, with a lower probability of damaging the biological sample under study. Quantum plasmonic biosensing relies on strong light/matter coupling and enables cavity sensing as well as surface-enhanced Raman scattering, which due to the strong interaction reveals a frequency range rich in Raman transitions that are barely visible in a standard frame. The contribution of microfluidics ensures a reduction in sample fluctuations and thus better control of the sample. Although this is a newcomer, there are already interesting results driving progress in quantum plasmonic biosensing, and the prospect that the next decade will play a crucial role in the solid establishment of this approach.

In conclusion, it should first be noted that these three buzzwords seem to tie together the world research of recent years, namely 'combined' as an indication of the integration of plasmonic techniques with other sensing techniques to increase sensitivity, 'wearable' because microfluidic is perfectly adaptable systems that detect, measure and track body fluids, and 'cost-effective' coherently with the intrinsic properties of microfluidics and lab-on-a-chip. By following the trend that has emerged in recent years, there is potential for the development of plasmonic biosensing in these directions.

**Disclosure statement.**

The author reports there are no competing interests to declare.

**Nomenclature**

| | |
|---|---|
| r, x,y,z space coordinates | c speed of light in vacuum |
| $k_{\parallel,in}(\omega)$ Incident wavevector | $E_i$ i=x,y,z component of the Electric field |
| $k_{\parallel,spp}(\omega)$ SPP wavevector | $E_{in}$ Incident electric field |
| $\varepsilon(\omega) = \varepsilon_m$ Complex permittivity of the metal | $E_r$ Reflected electric field |
| $\varepsilon_a$ Permittivity of analyte | $c = \omega/k \; ; \; \lambda = 2\pi c/\omega$ |
| $\varepsilon_d$ Permittivity of dielectric | $\varphi$ phase |
| $\varepsilon_0$ Dielectric constant in vacuum | $\kappa$ shape factor     in $E = E_0 \frac{(1+\kappa)\varepsilon_m}{\varepsilon_w + \kappa\varepsilon_m}$ |
| $\varepsilon_w$ Permittivity of waveguide | $\frac{2\alpha E_0}{4\pi r^3} + \frac{3\beta \dot{E}_0}{4\pi r^4 \varepsilon_0}$ $\alpha$ and $\beta$ are polarizability tensors of dipole and quadrupole |
| $k_{x,m} = \pm k_g$ In-plane wavevector of the order m of the grating | |
| $k_x$ x component of the wavevector | $\Delta\Phi$ phase uncertainty |
| k Wavevector | $\Delta I$ Amplitude uncertainty |
| $\Lambda$ Period of the grating | N Photon number |
| $n_p$ refractive index of prism | $|N\rangle$ Photon number state |
| $n_d$ dielectric refractive index | Q Quality factor |
| $n_m$ metal refractive index | $V/V_0$ Volume mode |
| $\vartheta$ angle of incidence | |
| $\omega$ light frequency | |
| $\lambda$ wavelength | |